\documentclass[aps,twocolumn,showpacs,superscriptaddress,groupedaddress]{revtex4}
\usepackage{dsfont,amsthm,amsbsy}
\usepackage{amssymb}
\usepackage{amsmath}
\usepackage{graphicx}
\usepackage{epstopdf}
\usepackage{subfigure}
\usepackage{natbib}
\usepackage{epsfig}
\usepackage{amsfonts}
\usepackage{mathrsfs}
\usepackage{sidecap}
\usepackage{lipsum}
\usepackage[toc,page,title,titletoc,header]{appendix}
\usepackage[colorlinks,linkcolor=blue,citecolor=blue,anchorcolor=blue]{hyperref}
\usepackage{dsfont,amsthm,amsbsy}
\usepackage{resizegather}
\usepackage{caption}
\usepackage{tikz}

\newcommand{\bl}{\begin{aligned}}
\newcommand{\el}{\end{aligned}}
\def\be{\begin{equation}}
\def\ee{\end{equation}}

\def\bi{\begin{itemize}}
\def\ei{\end{itemize}}
\def\bn{\begin{enumerate}}
\def\en{\end{enumerate}}
\def\bea{\begin{eqnarray}}
\def\eea{\end{eqnarray}}

\def\ba{\begin{array}}
\def\ea{\end{array}}
\def\bd{\begin{displaymath}}
\def\ed{\end{displaymath}}

\newcommand{\bS}{\bf S}
\begin{document}

\title{Magnetic quantum correlation in the 1D transverse-field XXZ model}

\author{Salimeh Mahdavifar}
\affiliation{Department of Physics, Alzahra University, 19834, Tehran,
Iran}

\author{Saeed Mahdavifar}
\affiliation{Department of Physics, University of Guilan, 41335-1914, Rasht, Iran}
\email[]{smahdavifar@gmail.com}

\author{R. Jafari}
\affiliation{Department of Physics, Institute for Advanced Studies in Basic Sciences (IASBS), Zanjan 45137-66731, Iran}
\affiliation{School of Physics, Institute for Research in Fundamental Sciences (IPM),
P.O.Box 19395-5531, Tehran, Iran}
\affiliation{Department of Physics, University of Gothenburg, SE 412 96 Gothenburg, Sweden}
\affiliation{Beijing Computational Science Research Center, Beijing 100094, China}
\email[]{rouhollah.jafari@physics.gu.se, rohollah.jafari@gmail.com}

\date{\today}
\begin{abstract}
One-dimensional spin-1/2 systems are well-known candidates to study the quantum correlations between particles. In the condensed matter physics, studies often are restricted to the 1st neighbor particles. In this work, we consider the 1D XXZ model in a transverse magnetic field (TF) which is not integrable except at specific points. Analytical expressions for quantum correlations (entanglement and quantum discord) between spin pairs at any distance are obtained for both zero and finite temperature, using an analytical approach  proposed by Caux \textit{et al.} [PRB \textbf{68}, 134431 (2003)].
We compare the efficiency of the QD with respect to the entanglement in the detection of critical points (CPs) as the neighboring spin
pairs go farther than the next nearest neighbors. In the absence of the TF and at zero temperature, we show that the QD for spin pairs farther than the 2nd neighbors is able to capture the critical points while the pairwise entanglement is absent.  In contrast to the pairwise entanglement, two-site quantum discord is effectively long-range in the critical regimes where it decays algebraically with the distance between pairs.
We also show that the thermal quantum discord between neighbor spins possesses strong distinctive
behavior at the critical point that can be seen at finite temperature and, therefore, spotlights the critical point
while the entanglement fails in this task.
\end{abstract}
\maketitle

\section{Introduction}\label{sec1}

Quantum phase transition (QPT) is one of the most interesting research topics in the condensed matter
physics. It is a phase transition theoretically occur at absolute zero temperature where the quantum
fluctuations play the dominant role \cite{Vojta2003}. Due to suppression of the thermal fluctuations
at zero temperature the ground state of the system is introduced as the representative of the system
where undergoes an abrupt change at the critical point (CP) \cite{Sachdev}. The ground state's wavefunction
of a many-body system near a CP at zero temperature is often nontrivial due to the long-range correlations among
the system's constituents. Quantum correlations could be responsible for these correlations \cite{Preskill} and
consequently, could be useful in studying the QPT. Entanglement, a type of quantum correlation was first pointed
out by Schordinger in 1935 \cite{schrodinger} as the characteristic feature of quantum mechanics. It has been widely
considered to be the main resource in most of the quantum information processing tasks \cite{Horodecki, Hill, Wootters, Modi2012}.
However, in the past few years, it has been known that, there exist quantum correlations which are not
spotlighted by entanglement measures. That is encompassed very efficiently in the formulation of so-called quantum discord (QD)
as a measure to represent the broadness of quantum correlations \cite{Ollivier, Oppenheim}. Moreover, there exist
intimations that the QD is the resource responsible for the speed up in deterministic quantum computation with one
quantum bit \cite{Datta2008, Lanyon}. Entanglement and QD have been studied extensively in a number of contexts, e.g.,
low dimensional spin models \cite{Dillenschneider08, sa22, Werlang10, Chen2010, Jafari2010, Tomasello2011, TOMASELLO2012}, open quantum
systems \cite{Jafari2015, Maziero2009, Shabani2009, Werlang2009, Modi2010}, biological \cite{Kamil2010}, and
relativistic \cite{Datta2009, Landulfo2010} systems.
Recently, pairwise QD and entanglement have been analyzed as a function of distance between spins in the
transverse field XY chain for both zero and finite temperatures cases \cite{Maziero10, Maziero2012, Tomasello2011, Campbell2013}.
It has been displayed that, at zero temperature, QD can capture a QPT even for situations where entanglement is absent.
Further, pairwise QD of two nearest neighbor spin in XXZ model can also indicate the critical points for finite temperatures
\cite{Werlang10, Tan2014}. Indeed, establishing finite quantum correlations between distant parties is undoubtedly imperative
to implement several quantum information processing tasks in many body systems with short-range interactions.
Along this direction, it has been shown that, QD length is enhanced by introducing disorder in a spin chain while,
entanglement length is not \cite{Sadhukhan2016}.

In this paper we study pairwise entanglement and QD at both zero and finite temperatures in the one dimensional
$XXZ$ chain in the presence of a transverse magnetic field which is not integrable except at specific points \cite{Andreas}.
Our motivation is related to very recent studies on 1D many-body quantum system of trapped ions \cite{Jurcevic14, Toskovic16}
where the entanglement between pairwise spins in a one-dimensional quantum system of trapped ions
has been observed \cite{Jurcevic14}. Moreover, it is experimentally reported that, constructed arrays of magnetic atoms on a
surface can be designed to behave like spin-1/2 XXZ Heisenberg chains in a TF \cite{Toskovic16}.
Consequently, the quantum correlation between different neighbor spins in the transverse field spin-1/2 XXZ Heisenberg chains
can be measured experimentally. Furthermore, in Ref. [\onlinecite{Glaser}] XXZ chains have been used to describe quantum
computers based on NMR and it can also be employed for solid state quantum computers \cite{Loss98, Loss99}.

The main aim of this study is to search the behavior of the entanglement and QD at zero and finite temperature for
spin pairs arbitrarily distant using an analytical approach by combining a Jordan-Wigner transformation with
a mean-field approximation \cite{Caux}. To the best of our knowledge, such contributions
have not been explored in previous works and can bring several new effects to the subject.
We show that the QD for spin pairs more distant than nearest-neighbors
is able to characterize QPT where pairwise entanglement is absent. This behavior is rather different from the behavior of
two-spin entanglement, which is typically short-range even in the critical regimes. Furthermore, an analysis displays that, the entanglement
and QD increase with magnetic field and temperature for certain regions of parameter space. We also show that the thermal QD is more
robust than the thermal entanglement as the distance between the spin pairs is increased.

Then, it is constructive to review the main features of the one dimensional XXZ chain in the presence and absence of a
transverse magnetic field.
The Hamiltonian of spin-1/2 XXZ Heisenberg chains in a TF is given by
%
%
\begin{eqnarray}
{\cal H} &=& J\sum_{n=1}^{N} [S^{x}_{n}
S^{x}_{n+1}+S^{y}_{n} S^{y}_{n+1}+\Delta S^{z}_{n} S^{z}_{n+1}],
\label{Hamiltonian}
\end{eqnarray}
%
%
where $S_{n}$ is the spin-1/2 operator of the $n$-th site. $J>0$ denotes the antiferromagnetic exchange coupling, $\Delta$ is the anisotropy parameter. The periodic boundary condition is considered. Known the ground state phase diagram of the XXZ at zero temperature \cite{Takahashi99}, the model
has three phases. In the limit $\Delta\gg 1$, the interactions in the XY plane will be ignored, thus the model
should be in an antiferromagnetic phase. On the other
hand, in the limit $\Delta \ll -1$, the model should be in a ferromagnetic phase.  In the intermediate region, the
system is in the gapless Luttinger liquid phase. These three phases are separated by two critical points (CPs). At $\Delta=1$, we have a continuous quantum phase transition (QPT) and, at $\Delta=-1$, we have a first-order transition. At zero temperature, the quantum correlation between the 1st neighbor spins in the XXZ model is studied \cite{Dillenschneider08}. It infers that the quantum correlation between the 1st neighbor spins is maximal at the critical point $\Delta=1$ \cite{Cai06}. The critical point $\Delta=-1$ is not conformal and it has recently attracted some attention \cite{Banchi09, Alba13, Stasinska14}. It is shown that the finite-size corrections to the energy per site non trivially vanish in the ferromagnetic $\Delta\rightarrow -1^{+}$ isotropic limit. The multi-partite quantum nonlocality is also investigated in this model\cite{Sun14}. At finite temperature, the quantum correlations between the 1st neighbor spins are also investigated in this model \cite{Werlang10, Werlang11}. It is inferred that the quantum phase transitions have a decisive influence on a system's physical property  not only for small temperatures, but also for enough above temperatures where quantum fluctuations no longer dominate.

One of the striking effects is the dependence of the physical properties of the 1D spin-1/2 XXZ model on the direction of the applied magnetic field. It is known that adding a transverse magnetic field ($h \sum_{n=1}^{N} S^{x}_{n}$) to the XXZ model breaks the $U(1)$ symmetry and the exact integrability  is  lost \cite{Kurmann82, Dmitriev02, Mahdavifar06}.  The TF induces a gap in the region $-1 <\Delta \leq 1$ and the ground state has the long-range spin-flop order up to a critical TF. In the region $\Delta>1$ ($\Delta\leq -1$), by applying the TF, a phase transition from the Neel (ferromagnetic) phase to a phase with saturated magnetization along field occurs at a critical TF.  Moreover, a completely factorized \cite{Giampaolo08}  ground state may occur at a specific value of the TF, $h_f=J\sqrt{2 (1+\Delta)}$. It was shown that the entanglement of the factorized state in a TF is remarkably singled out by entanglement \cite{Roscilde04, Abouie10, Amico12}. The experimental observations on the quasi-one dimensional spin-1/2 antiferromagnet $Cs_{2}CoCl_{4}$ are a realization of the effect of such a TF on the low energy behavior of a 1D XXZ model \cite{Kenzelmann02, Breunig13}.

The paper is organized as follows. In the next section, we introduce the model and express an analytical form for the entanglement and the QD. In section \ref{sec3}, analytical results will be presented. Finally, we conclude and summarize our results in section \ref{sec4}.

\section{Quantum correlations}\label{sec2}

Initially, by performing a rotation of spins around the $y$ axis by $\pi/2$, the 1D spin-1/2 XXZ in a transverse field is transformed as\cite{Dmitriev02}
\begin{eqnarray}
{\cal H} &=& \sum_{n=1}^{N} [J \Delta S^{x}_{n} S^{x}_{n+1}+J (S^{y}_{n}
S^{y}_{n+1}+  S^{z}_{n} S^{z}_{n+1})]\nonumber \\
&-& h \sum_{n=1}^{N}S^{z}_{n}.
\label{Hamiltonian}
\end{eqnarray}
At second, by applying the Jordan-Wigner transformation
\begin{eqnarray}
S^{+}_{n}&=& a_{n}^{\dag}(e^{i\pi \sum_{l<n} a_{l}^{\dag}a_{l}}),~~
S^{-}_{n}= (e^{-i\pi \sum_{l<n} a_{l}^{\dag}a_{l}})a_{n},\\
S^{z}_{n}&=& a_{n}^{\dag}a_{n}-\frac{1}{2},
\label{fermion operators}
\end{eqnarray}
the Hamiltonian is mapped onto a Hamiltonian of 1D interacting fermionic
system
\begin{eqnarray}
{\cal H}&=& \frac{J(\Delta-1)}{4} \sum_{n} (a^{\dag}_{n}a^{\dag}_{n+1}+h.c.)\nonumber \\
&+&\frac{J(\Delta+1)}{4} \sum_{n} (a^{\dag}_{n}a_{n+1}+h.c.)\nonumber \\
&+& J \Delta \sum_{n} [a^{\dag}_{n}a_{n} (a^{\dag}_{n+1}a_{n+1}-1)]\nonumber \\
&-& h \sum_{n} a^{\dag}_{n}a_{n}.
\label{fermionic Hamiltonian}
\end{eqnarray}
At third, using the Wick's theorem, the fermion interaction term is decomposed by some order parameters which are
related to the two-point correlation functions as
\begin{eqnarray}
\gamma_1&=&\langle a^{\dag}_{n}a_{n} \rangle , \nonumber \\
\gamma_2&=&\langle a^{\dag}_{n}a_{n+1} \rangle ,\nonumber \\
\gamma_3&=&\langle a^{\dag}_{n}a^{\dag}_{n+1}\rangle .
\end{eqnarray}
By utilizing these order parameters and performing a Fourier transformation as $a_{n} = \frac{1}{\sqrt{N}} \sum _{k} e^{-ikn} a_{k}$, and also Bogoliobov transformation
\begin{eqnarray}
a_{k}=cos(k) \alpha_k -i~ sin(k) \alpha^{\dag}_{-k},
\end{eqnarray}
the diagonalized Hamiltonian is given by
\begin{eqnarray}
{\cal H}_{f}=\sum_{k=-\pi}^{\pi}\varepsilon(k) (\alpha_{k}^{\dagger} \alpha_{k}-\frac{1}{2}).
\label{Hamiltonian d}
\end{eqnarray}
Where the energy spectrum is
\begin{eqnarray}
\varepsilon(k) &=&  \sqrt{a(k)^2+b(k)^2}, \nonumber\\
a(k)&=&(\frac{J(\Delta+1)}{2}-2 \gamma_2) \cos(k)+(2 \gamma_1 -1) J-h, \nonumber\\
b(k)&=&(2 J \gamma_3+\frac{J(\Delta-1)}{2}) \sin(k).
\end{eqnarray}
One should note that the following equations should be satisfied self-consistently
\begin{eqnarray}
\gamma_1&=&\frac{1}{2}-\frac{1}{\pi} \int_{0}^{\pi} dk~\frac{a(k)}{\varepsilon(k)}(\frac{1}{2}-f(k)),  \nonumber  \\
\gamma_2&=&-\frac{1}{\pi} \int_{0}^{\pi} dk~ \cos(k) \frac{a(k)}{\varepsilon(k)} (\frac{1}{2}-f(k)), \nonumber \\
\gamma_3&=&-\frac{1}{2 \pi} \int_{0}^{\pi} dk~ \sin(k) \frac{b(k)}{\varepsilon(k)}(\frac{1}{2}-f(k)),
\label{self}
\end{eqnarray}
where the Fermi distribution function is $f(k)=\frac{1}{1+e^{-\beta \varepsilon(k)}}$, $\beta=\frac{1}{K_{B}T}$ and the Boltzmann constant is taken $K_B=1$.
The concurrence between two spins at site $i$ and $j$ can be achieved from
the corresponding reduced density matrix $\rho_{ij}$, which in
the standard basis is expressed as
\begin{eqnarray}
\rho_{i,j}= \left(
             \begin{array}{cccc}
               <P_{i}^{\uparrow}P_{j}^{\uparrow}> & <P_{i}^{\uparrow}{\bS}_{j}^{-}> & <{\bS}_{i}^{-}P_{j}^{\uparrow}> & <{\bS}_{i}^{-}{\bS}_{j}^{-}> \\
               <P_{i}^{\uparrow}{\bS}_{j}^{+}> & <P_{i}^{\uparrow}P_{j}^{\downarrow}> & <{\bS}_{i}^{-}{\bS}_{j}^{+}> & <{\bS}_{i}^{-}P_{j}^{\downarrow}>\\ <{\bS}_{i}^{+}P_{j}^{\uparrow}> & <{\bS}_{i}^{+}{\bS}_{j}^{-}> & <P_{i}^{\downarrow}P_{j}^{\uparrow}> & <P_{i}^{\downarrow}{\bS}_{j}^{-}>\\
               <{\bS}_{i}^{+}{\bS}_{j}^{+}> & <{\bS}_{i}^{+}P_{j}^{\downarrow}> & <P_{i}^{\downarrow}{\bS}_{j}^{+}> & <P_{i}^{\downarrow}P_{j}^{\downarrow}>\\
               \end{array}\nonumber
               \right),
\label{density matrix1}
\end{eqnarray}
where $P^{\uparrow}=\frac{1}{2}+S^{z}, P^{\downarrow}=\frac{1}{2}-S^{z}$. The brackets symbolize the thermodynamic average values at zero and finite temperature. In this literature, we introduce $S^{\pm}$ as $S^{\pm}= S^{x}\pm i S^{y}$. By applying the Jordan-Wigner transformation, the reduced density matrix will be written as\cite{Gong09, Mehran14, Soltani14, Khastehdel16}
\[
\rho_{i,j} =
\left( {\begin{array}{cccc}
  X^{+}  & 0 &  0 & 0\\
    0 &  Y^{+}  & Z^{*} & 0\\
 0 & Z & Y^{-}  & 0\\
 0 & 0 & 0 & X^{-}
 \end{array} } \right),
\]
where $X^{+}=\langle n_{i} n_{j}\rangle~(n_j=a_{j}^{\dagger}a_{j})$,  $Y^{+}=\langle n_i(1-n_{j})\rangle$, $Y^{-}=\langle n_{j}(1-n_{i})\rangle$, $Z=\langle a_{i}^{\dagger}a_{j} \rangle$ and $X^{-}=\langle 1-n_i- n_{j}+n_i n_{j}\rangle$. Thus, the concurrence is transformed into
\begin{eqnarray}
\nonumber
C_{i,j}=\mbox{max}\{0,2 (|Z|-\sqrt{X^{+} X^{-}})\}.
\label{concurrence}
\end{eqnarray}
%
\begin{figure*}
\centerline{
\includegraphics[width=0.45\linewidth]{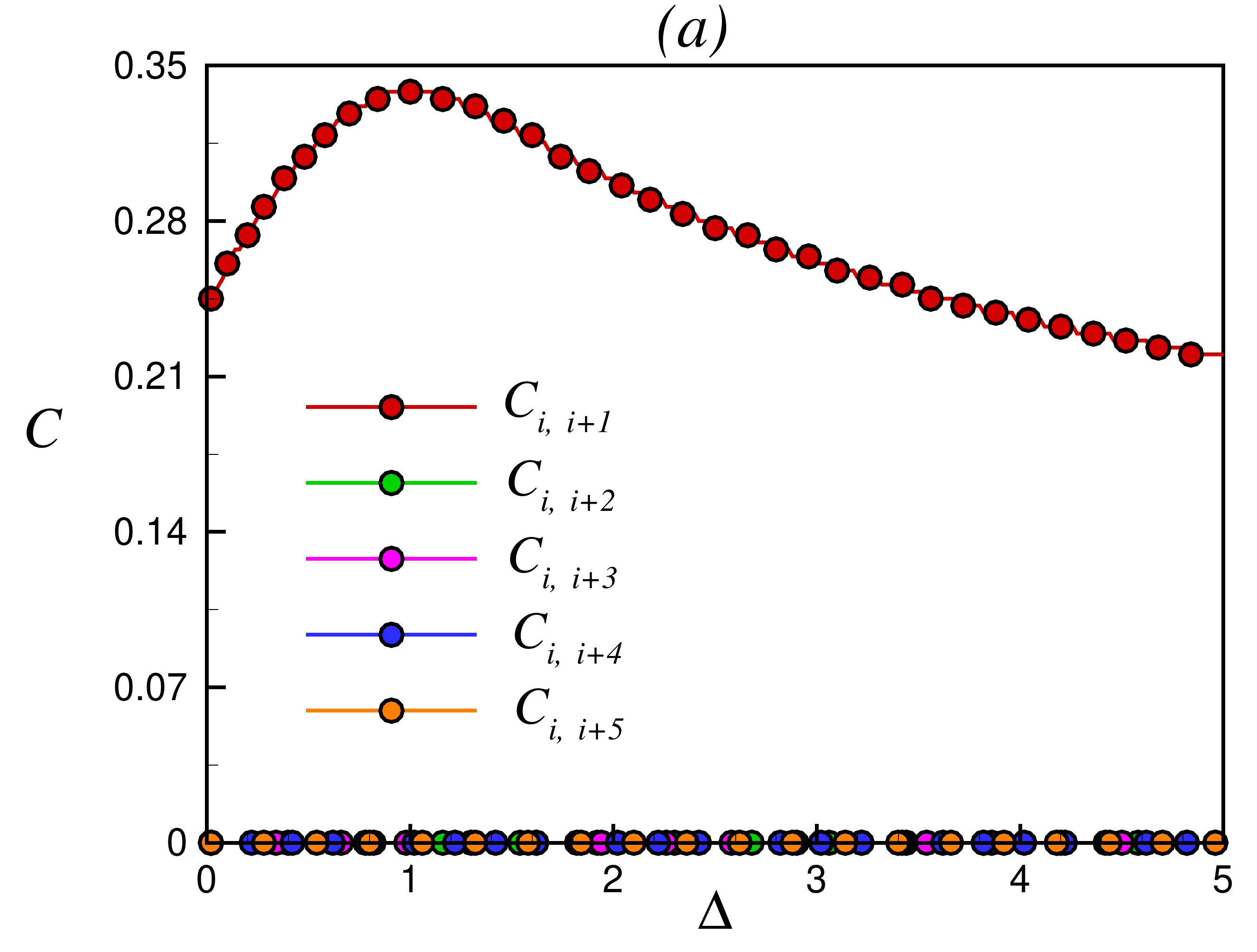}
\hspace{1cm}
\includegraphics[width=0.45\linewidth]{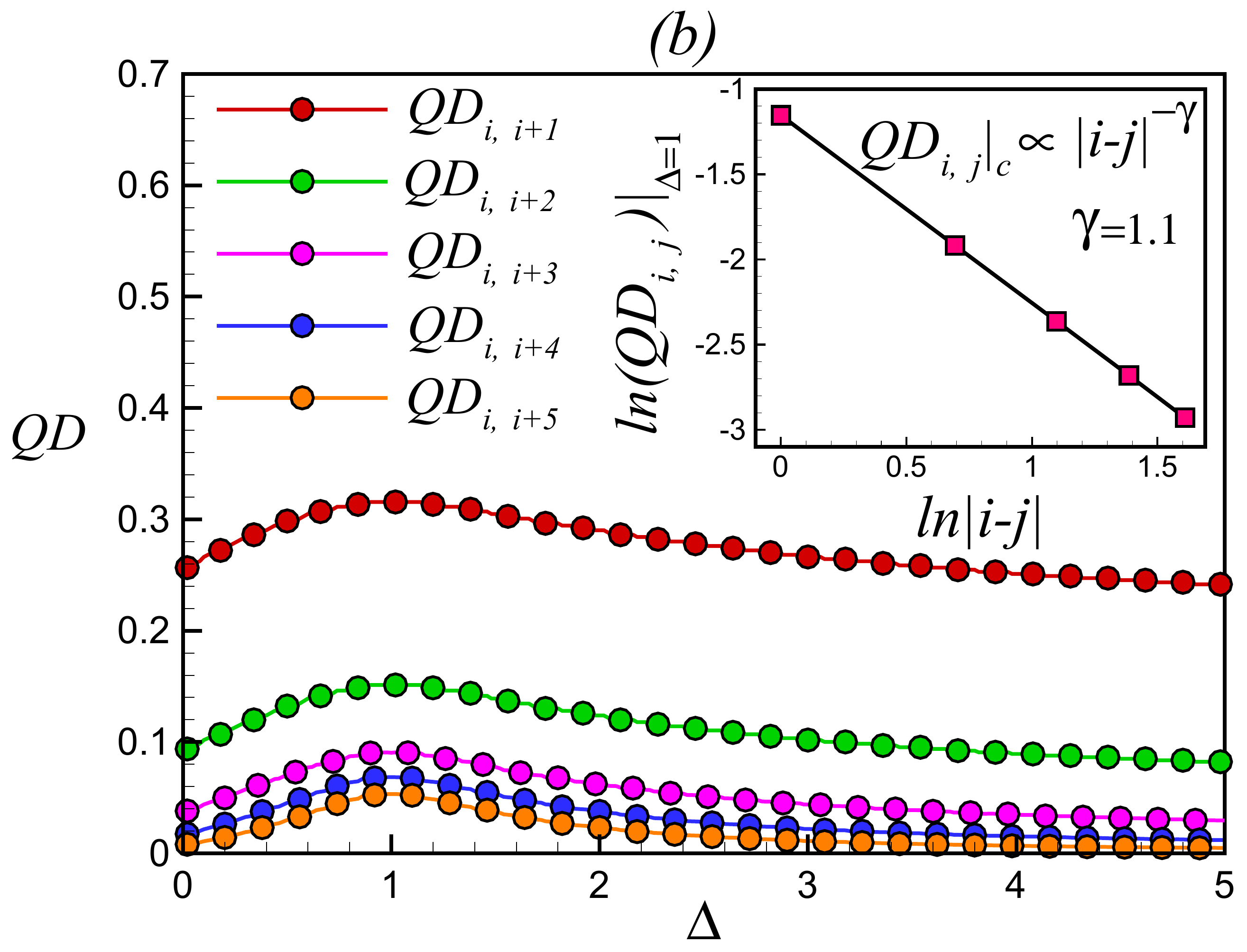}}
\caption{(Color online.) (a) Entanglement of formation and (b) quantum
discord  between  the first, second, third, fourth and fifth nearest neighbors as
a function of anisotropy at zero temperature and zero magnetic field.
Inset: Scaling behaviour of quantum discord at the critical point $\Delta=1$ in terms
of distances between spins pair.}
\label{Fig1}
\end{figure*}
Now, to study the quantum discord (QD), we follow the Sarandy's prescription \cite{sa22}. The mutual information is given as
\begin{eqnarray}
\mathcal{I}(\rho_{i,j})=S(\rho_i)+S(\rho_{j})+\sum_{\alpha=0}^{3}\lambda_{\alpha}\log\lambda_{\alpha},
\label{eq28}
\end{eqnarray}
where
\begin{eqnarray}
S(\rho_i)&=&S(\rho_{j})=\\
&-&\Big[(\frac{1+c_3}{2})\log(\frac{1+c_3}{2})+(\frac{1-c_3}{2})\log(\frac{1-c_3}{2})\Big],\nonumber
\label{eq29}
\end{eqnarray}
$\lambda_{\alpha}$ is the eigenvalue of $\rho_{i,j}$ and new variables are related to the elements of the density matrix as
\begin{eqnarray}
c_1&=&2 Z,~~c_2=X^{+}+X^{-}-Y^{+}-Y^{-},\nonumber\\
c_3&=&X^{+}-X^{-},
\label{eq26}
\end{eqnarray}
To investigate the classical correlations between pair spins located at sites $i$ and $j$, one should introduce
a set of projectors for a local measurement on part $(j)=B$ given by
$\{B_{k'}=V\Pi_{k'}V^\dag\}$ where $\{\Pi_{k'}=|k'\rangle \langle k'|:k'=0,1\}$ is the set of projectors
on the computational basis $|0\rangle\equiv|\uparrow\rangle$ and $|1\rangle\equiv|\downarrow\rangle$
and $V \in U(2)$. $V$ is parametrized as
\begin{equation}
V=
\begin{pmatrix}
\cos \frac{\theta}{2} & \sin \frac{\theta}{2}e^{-i\phi} \\
\sin \frac{\theta}{2}e^{i\phi} & -\cos \frac{\theta}{2} \\
\end{pmatrix},
\label{eq30}
\end{equation}
where $0\leq\theta\leq\pi$ and $0\leq\phi<2\pi$ and they can be interpreted as the azimuthal and polar angles of a qubit over the Bloch sphere. After the measurement ${B_{k'}}$, the physical state of the system  will change
to one of the following states
\begin{eqnarray}
\label{eq32}
\rho_0&=&\left(\frac{I}{2}+\sum_{j=1}^{3}q_{0j}S_{j}\right)\otimes(V\Pi_0V^\dag),\\
\label{eq33}
\rho_1&=&\left(\frac{I}{2}+\sum_{j=1}^{3}q_{1j}S_{j}\right)\otimes(V\Pi_1V^\dag),
\end{eqnarray}
where
\begin{eqnarray}
q_{k'1}&=&(-1)^{k'} c_1\left[\frac{\sin\theta\cos\phi}{1+(-1)^{k'}c_3\cos\theta}\right],\nonumber\\
q_{k'2}&=&\tan\phi q_{k'1},\nonumber\\
q_{k'3}&=&(-1)^{k'}\left[\frac{c_2\cos\theta+(-1)^{k'}c_3}{1+(-1)^{k'}c_3\cos\theta}\right].
\label{eq34}
\end{eqnarray}
Then, by evaluating the von Neumann entropy from
Eqs.~(\ref{eq32}) and (\ref{eq33}) and using that $S(V\Pi_0V^\dag)=0$, we obtain
{\small
\begin{eqnarray}
S(\rho_{k'})=-(\frac{1+\theta_{k'}}{2})\log(\frac{1+\theta_{k'}}{2})+(\frac{1-\theta_{k'}}{2})\log(\frac{1-\theta_{k'}}{2}),
\nonumber\\
\label{eq35}
\end{eqnarray}
}
with
$\theta_{k'}=\sqrt{\sum_{j=1}^{3}q^{2}_{{k'}j}}$.
Finally, the classical correlation for the spin pair will be given by
\begin{small}
\begin{eqnarray}
\mathcal{C}\rho_{i,j})=
\max_{\{\Pi_i^B\}} \left(S(\rho_i)-\frac{S(\rho_0)+S(\rho_1)}{2}-c_3\cos\theta\frac{S(\rho_0)-S(\rho_1)}{2} \right),\nonumber
\label{eq37}
\end{eqnarray}
\end{small}
and the QD is determined as
\begin{eqnarray}
QD=\mathcal{I}(\rho_{i,j})-\mathcal{C(}\rho_{i,j}).
\label{QD}
\end{eqnarray}
%
\begin{figure*}[t]
\centerline{
\includegraphics[width=0.32\linewidth]{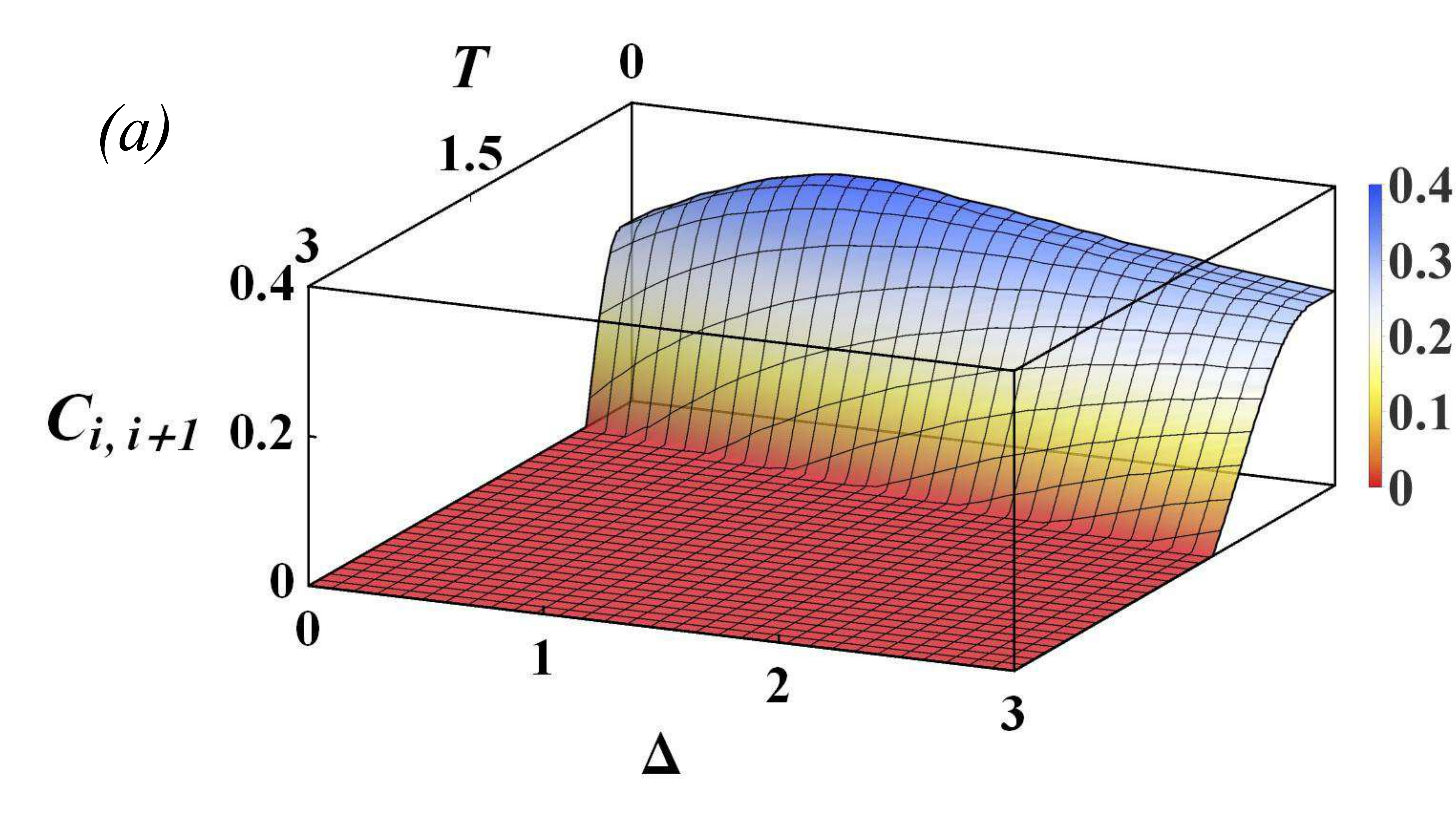}
\includegraphics[width=0.34\linewidth,height=0.18\linewidth]{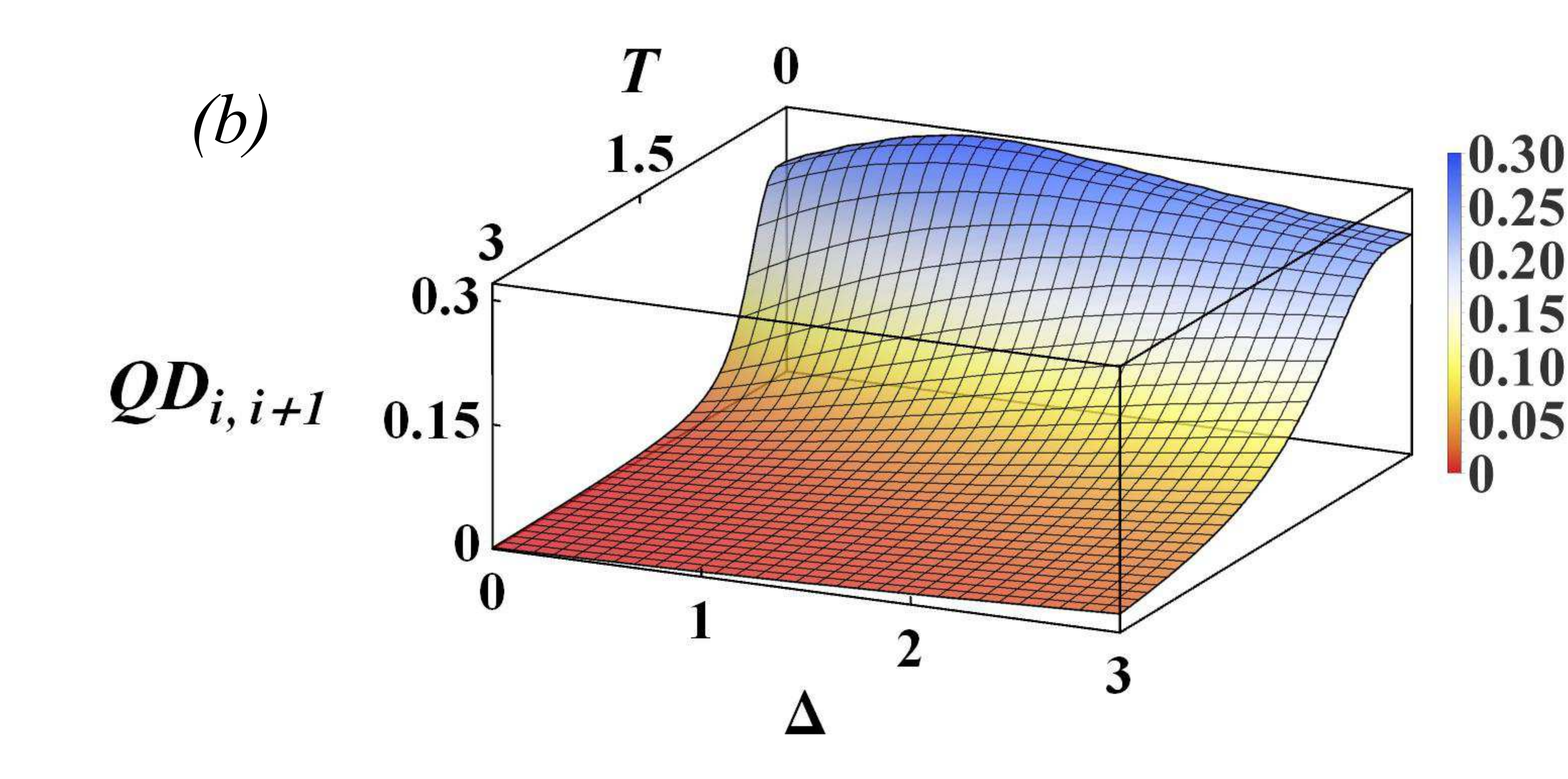}
\includegraphics[width=0.34\linewidth,height=0.18\linewidth]{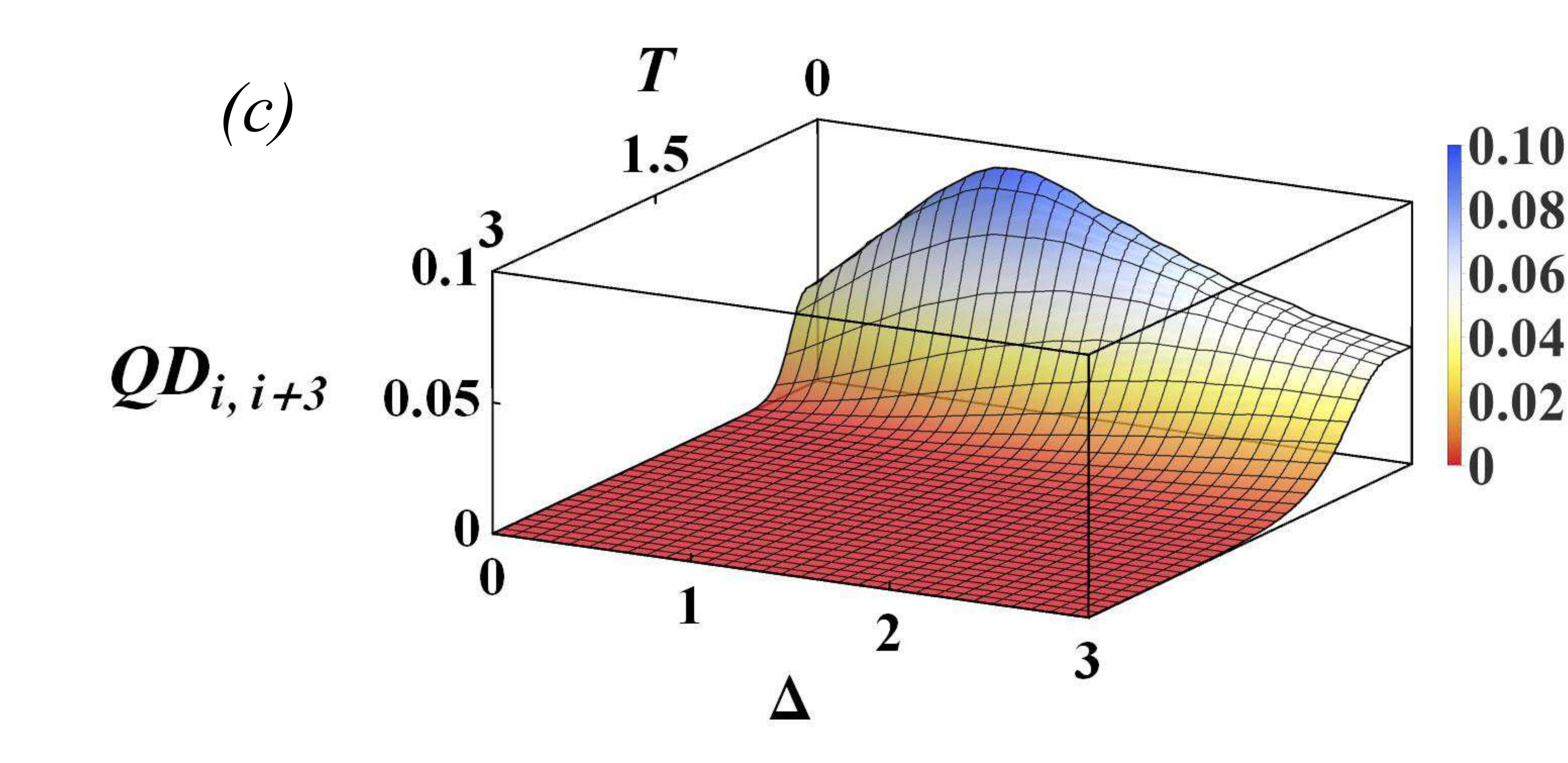}}
\caption{(Color online.) The three-dimensional panorama of Thermal (a) entanglement between first nearest neighbors,
(b) quantum discord between first nearest neighbors and (c) quantum discord between second nearest neighbor
spin pairs at zero magnetic field.}
\label{Fig2}
\end{figure*}
%
%
%
\begin{figure*}[t]
\centerline{
\includegraphics[width=0.36\linewidth]{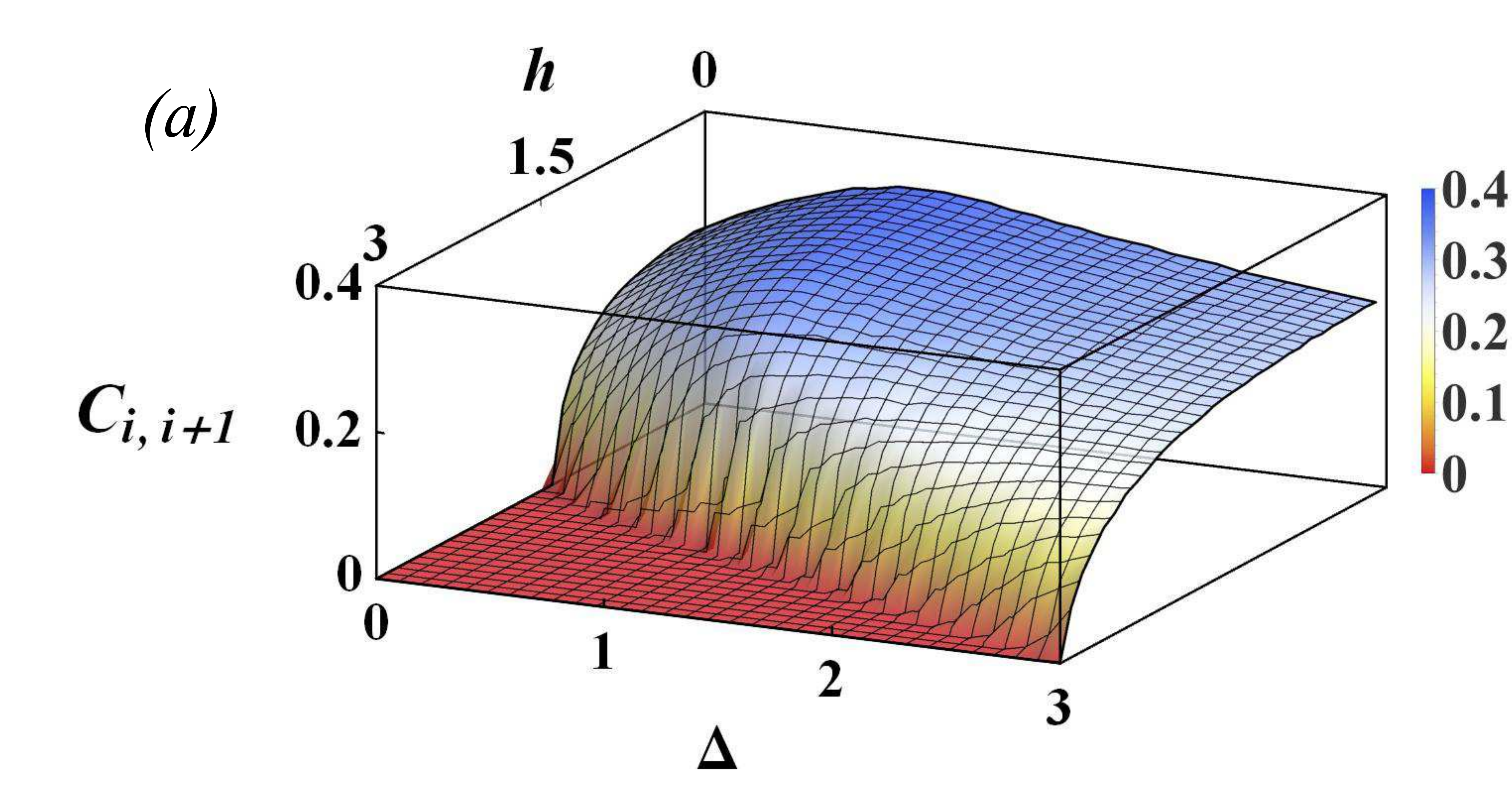}
\includegraphics[width=0.36\linewidth]{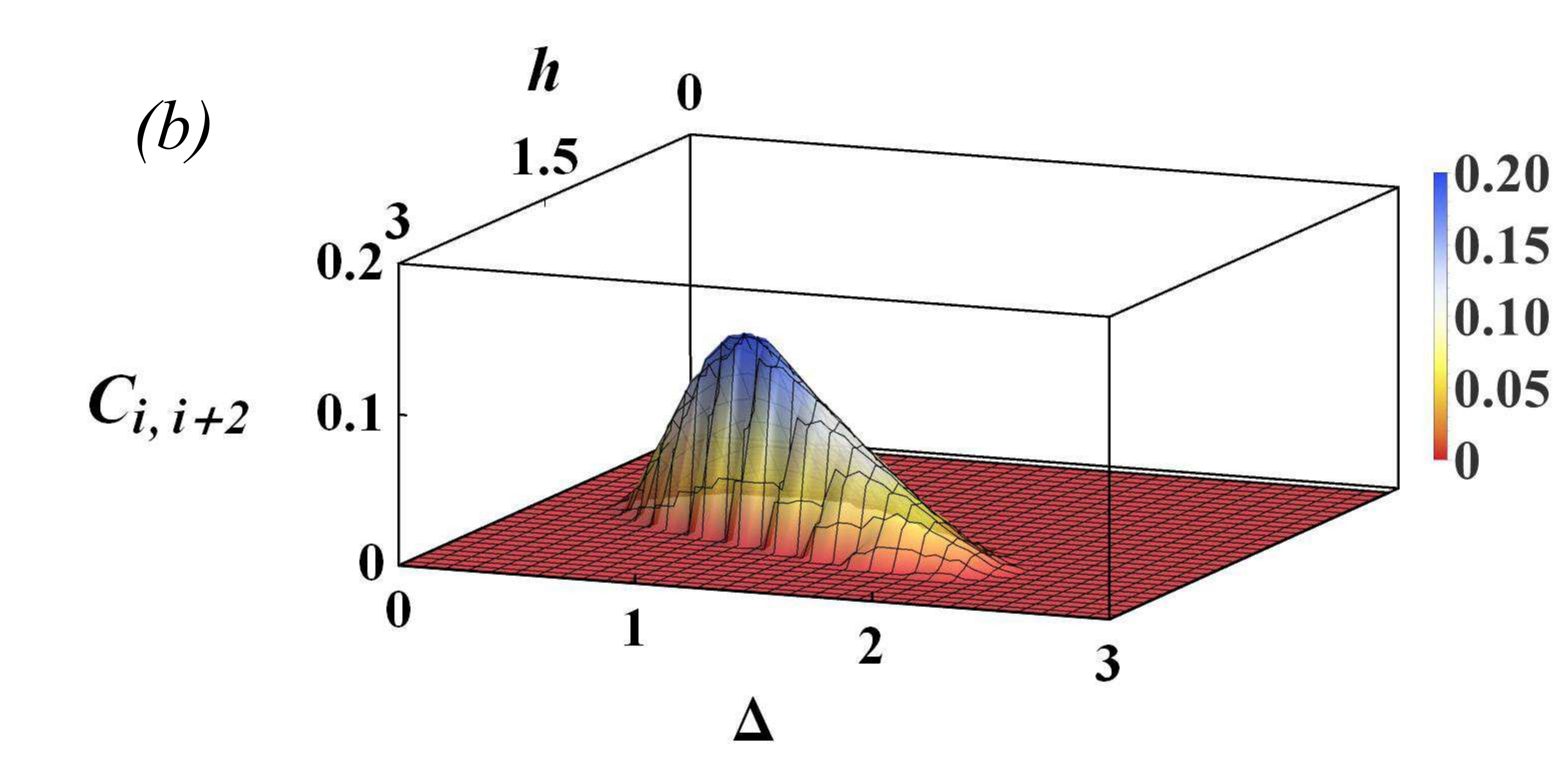}
\includegraphics[width=0.27\linewidth]{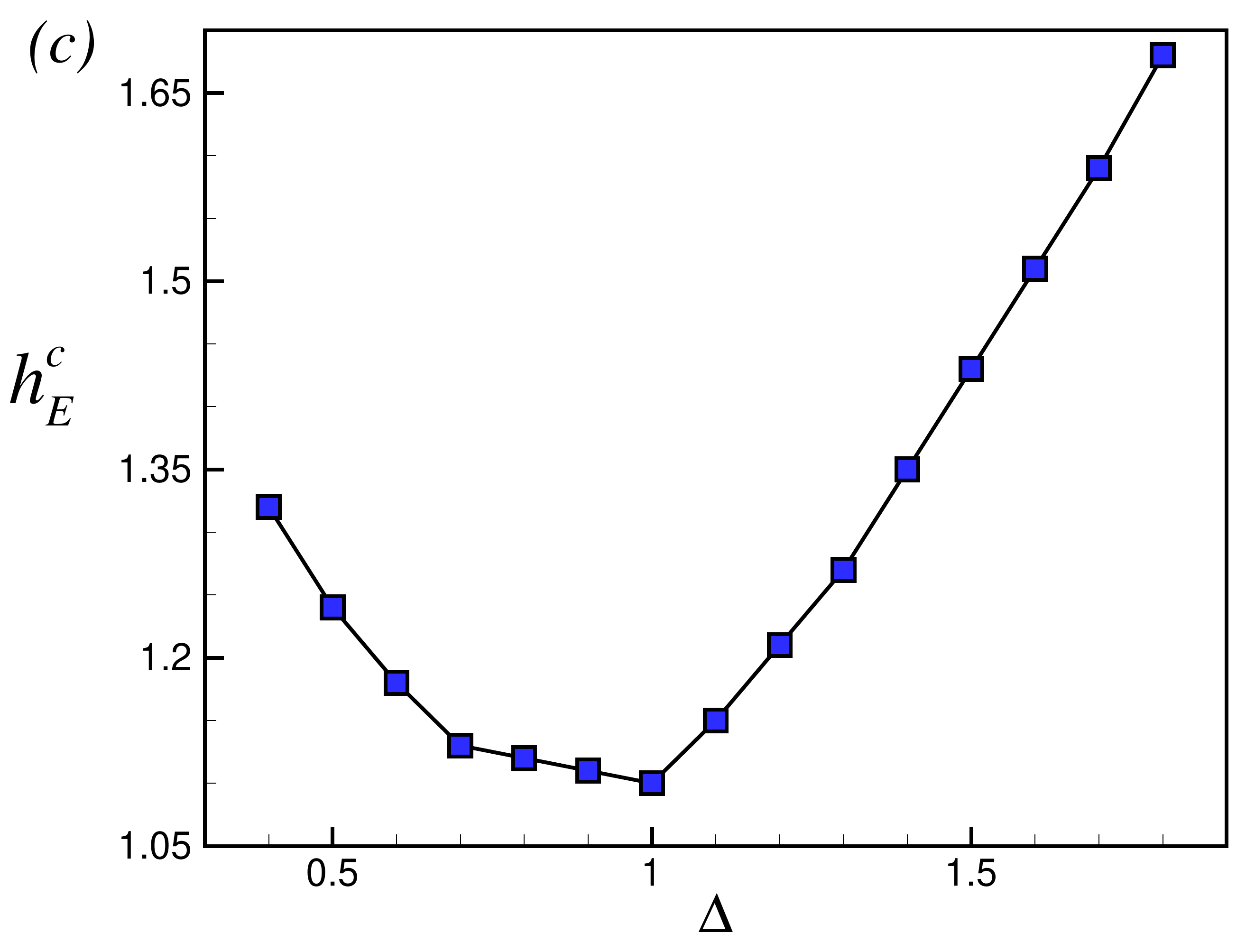}}
\caption{(Color online.) Three-dimension of entanglement as a function
of magnetic field and anisotropy $\Delta$ at zero temperature, between the
(a) first (b) second nearest neighbors spins. (c) The critical entangled-field
as a function of anisotropy parameter $\Delta$, at $T=0$.}
\label{Fig3}
\end{figure*}
%
%
\begin{figure*}[t]
\centerline{
\includegraphics[width=0.33\linewidth]{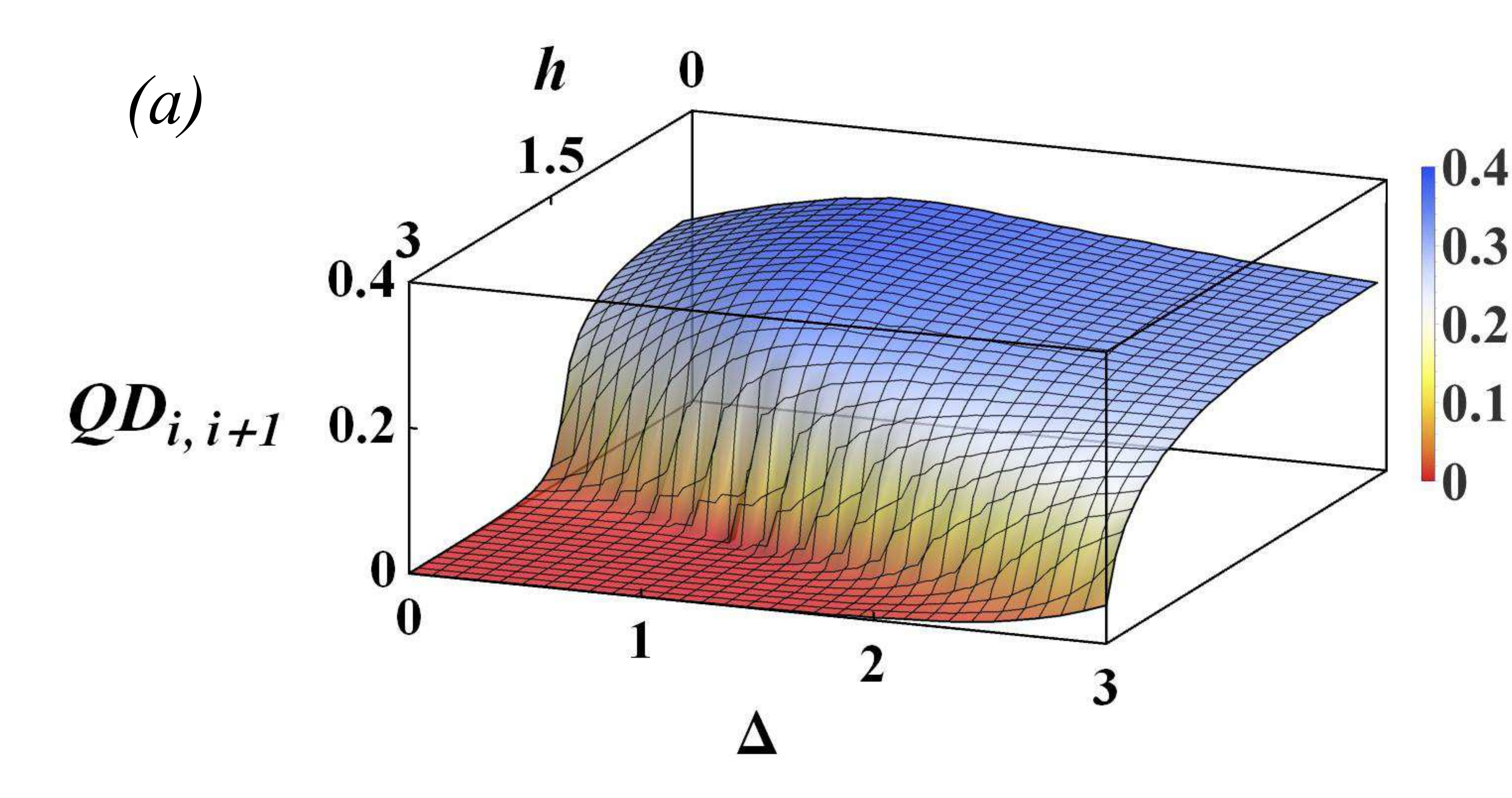}
\includegraphics[width=0.33\linewidth]{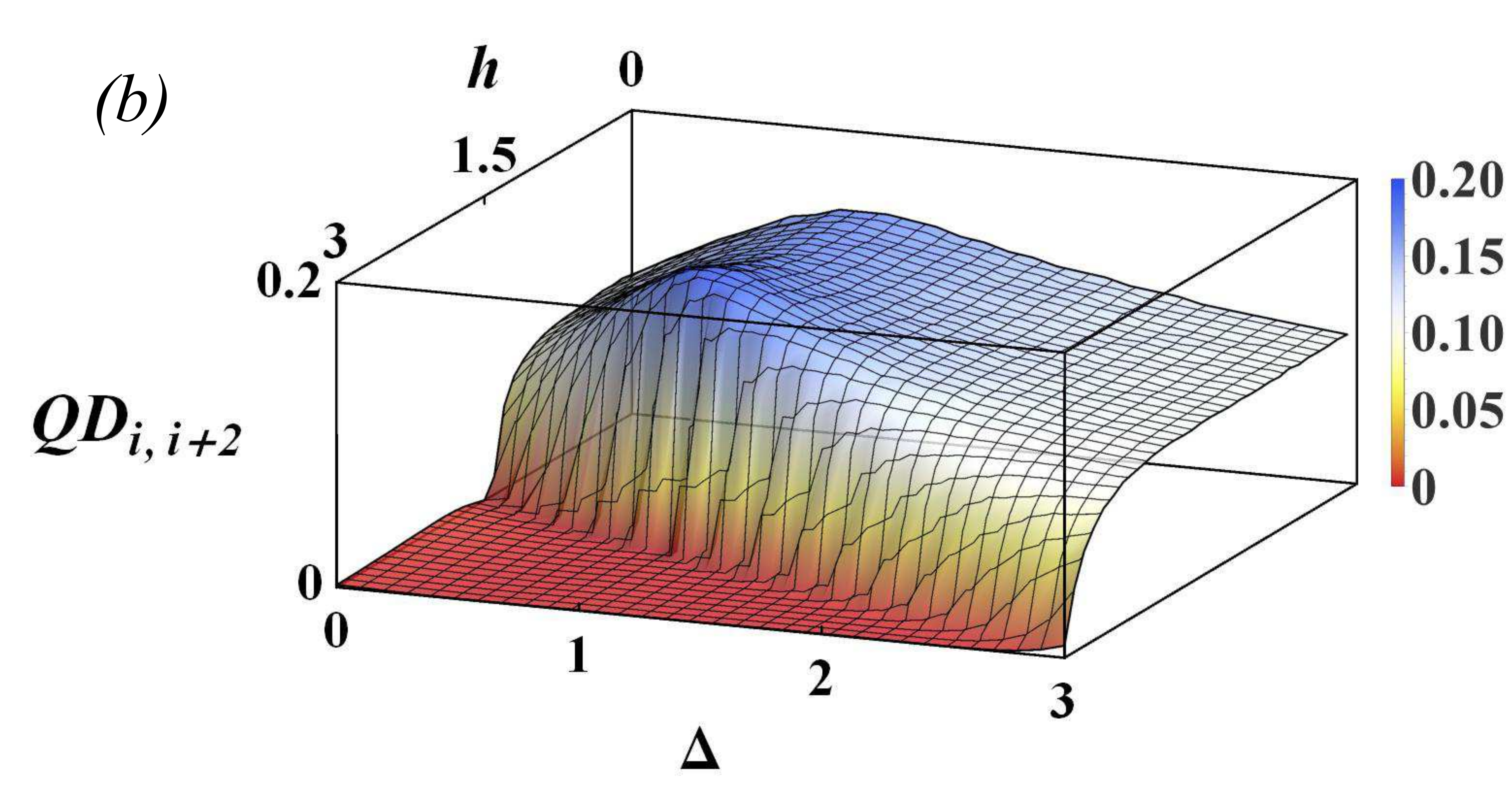}
\includegraphics[width=0.33\linewidth]{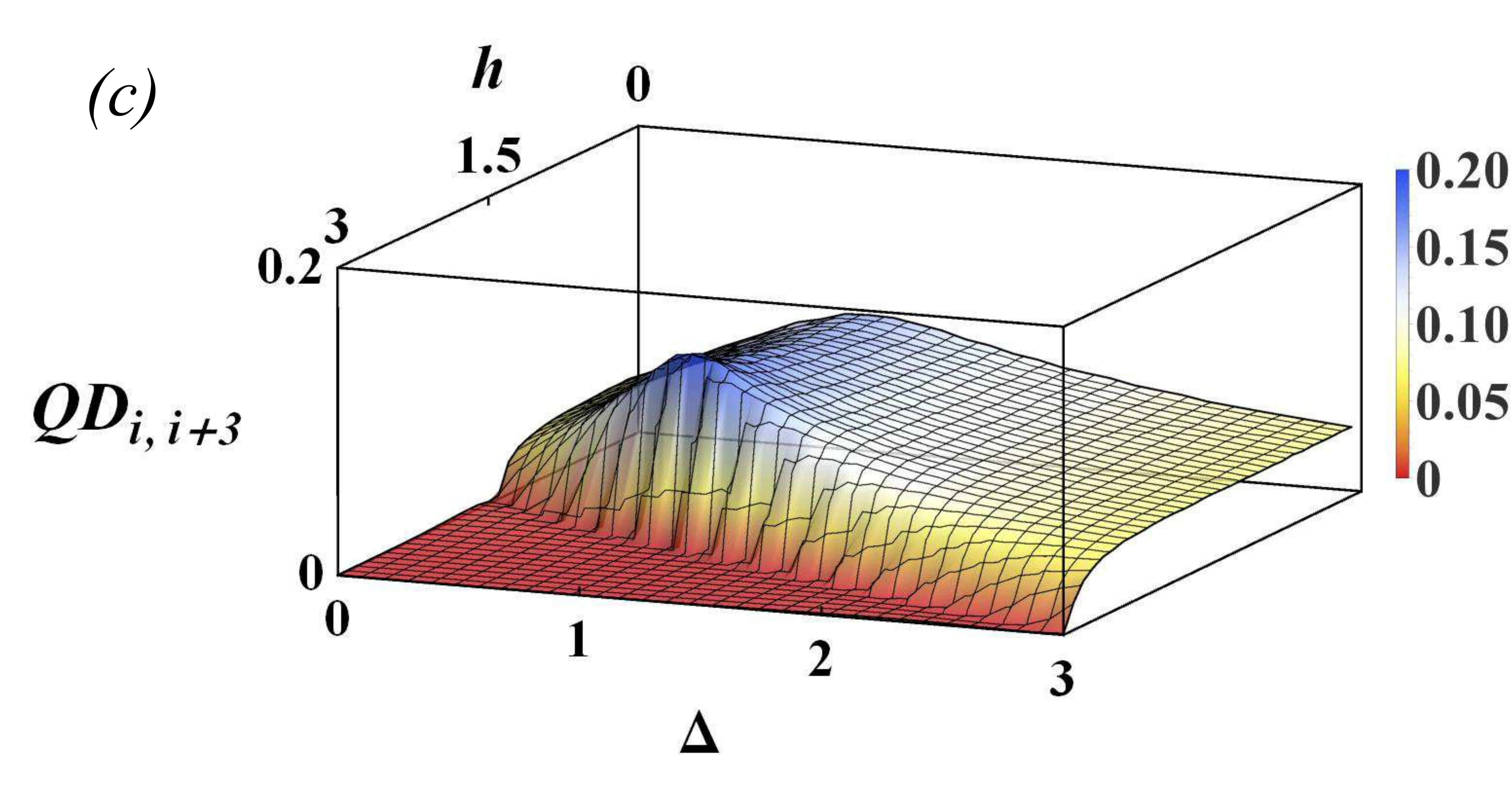}}
\caption{(Color online.) Three-dimension view of quantum discord as a function
of magnetic field and anisotropy $\Delta$ at zero temperature, between the
(a) first (b) second and (c) third nearest neighbors spins.}
\label{Fig4}
\end{figure*}
%
\section{Results}\label{sec3}

In this section, we report our numerical simulations results, based on the analytical approach,
of the entanglement and the QD between spin pairs.

We begin our analysis by studying the behavior of entanglement and QD as a function of the anisotropy parameter in
the absence of the TF at zero temperature where the model is integrable \cite{Andreas}.
QCs between the 1st, 2nd, 3rd, 4th and 5th neighbor spins have been depicted versus $\Delta$ in
Fig. \ref{Fig1}(a)-(b).
As seen in Fig. \ref{Fig1}(a), only the 1st neighbor spins are entangled in the whole range of
the anisotropy parameter $\Delta\geq0$ and is maximal at the critical point $\Delta_c=1.0$.
One can clearly see in Fig. \ref{Fig1}(b) that, the QD of  the 1st, 2nd, 3rd, 4th and 5th
neighboring pairs is nonzero and as expected, it decreases by increasing the distance between spin pairs .
Also, the QD of all the 1st, 2nd, 3rd, 4th and 5th neighboring pairs reaches to its maximum at the
critical point $\Delta_c=1.0$. A more detailed analysis shows that the QD of spin pairs at the critical point
decays algebraically with the distance between pairs $QD_{i, j}|_{c}\propto |i-j|^{-\gamma}$ with
$\gamma=-1.1$ (inset Fig. \ref{Fig1}(b)). This behavior reveals that, there exist long-range quantum correlations
as quantified by quantum discord which decay as a function of distance in short-range magnetic interaction system.
This is in contrast with the short-ranged behavior of the pairwise entanglement.

In order to show that, whether the thermal QD (TQD) of spin pairs is able to pinpoint the critical point $\Delta_{c}=1$
at finite temperature, we have plotted the thermal entanglement and TQD versus the anisotropy and temperature
in Figs. \ref{Fig2}(a)-(c) for $h=0$. The analysis shows that, while the maximum of the low temperature thermal entanglement
does not occur at the critical point $\Delta_{c}=1$, the phase transition point can be
signaled by the maximum of the TQD of the spin pairs at low temperature even for spin pairs more distant
than nearest-neighbors (Fig. \ref{Fig2}(c)). The maximum value is the result of an optimal mixing of all eigenstates
in the system. Although the maximum value of the low temperature TQD decreases as the distance between
the spin pairs increases, the slope in the critical region gets more visible for far neighbors (Fig. \ref{Fig2}(c)).
So, the low temperature TQD between far neighbors can be used to characterize the zero temperature phase transition.
These results are qualitatively in agreement with the results of Ref. [\onlinecite{Werlang10}]
where the exact solution of the model is presented by solving a set of non-linear integral equations.

In addition, at high temperature, both entanglement and QD between the 1st neighbor spins reduce by increasing the temperature.
As a result, the entanglement and the TQD become zero at the critical temperatures $T_{c}^{E}(\Delta)$ and $T_{c}^{D}(\Delta)$
respectively. More analysis shows that the critical temperatures, $T_{c}^{E, D}(\Delta)$ where quantum-classical
phase transition occurs, decrease by increasing the distance between spin pairs and increase by enhancing the anisotropy
parameter $\Delta$. Moreover, an increment of temperature decreases the TQD between arbitrary distant spin pairs and
can not create entanglement between spin pairs farther than first nearest neighbors.

The next step is to examine how the entanglement and QD capture the QPT in the presence of a transverse field at both zero
and finite temperature. To this end, we have calculated the entanglement and QD as a function of the TF and anisotropy
at zero temperature and $T=0.2$. Three dimensional view of quantum entanglement between the 1st and 2n neighbor spins has been
depicted in Figs. \ref{Fig3}(a)-(b) versus $h$ and $\Delta$ for $T=0$.
%
\begin{figure*}
\centerline{
\includegraphics[width=0.33\linewidth]{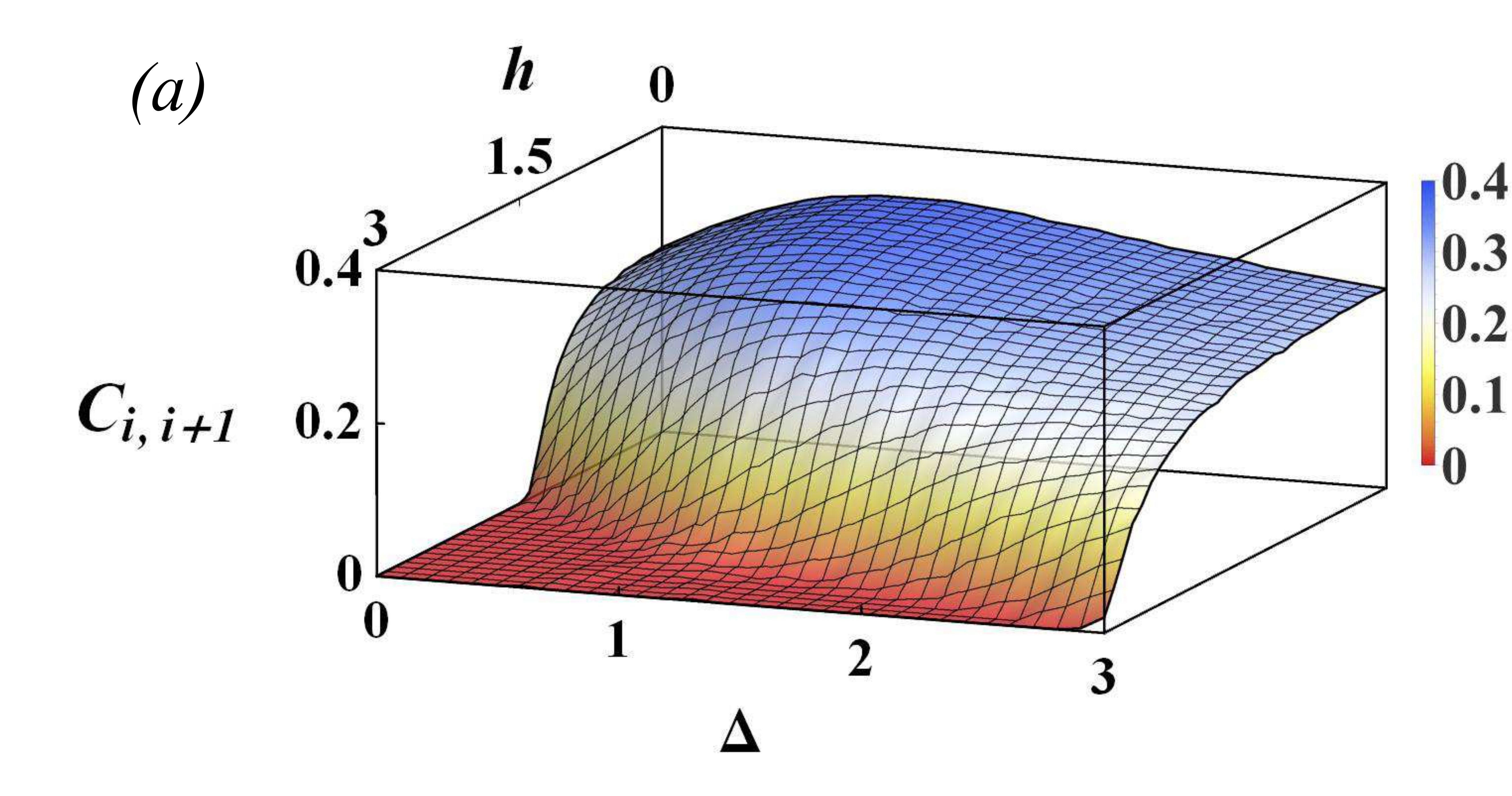}
\includegraphics[width=0.33\linewidth]{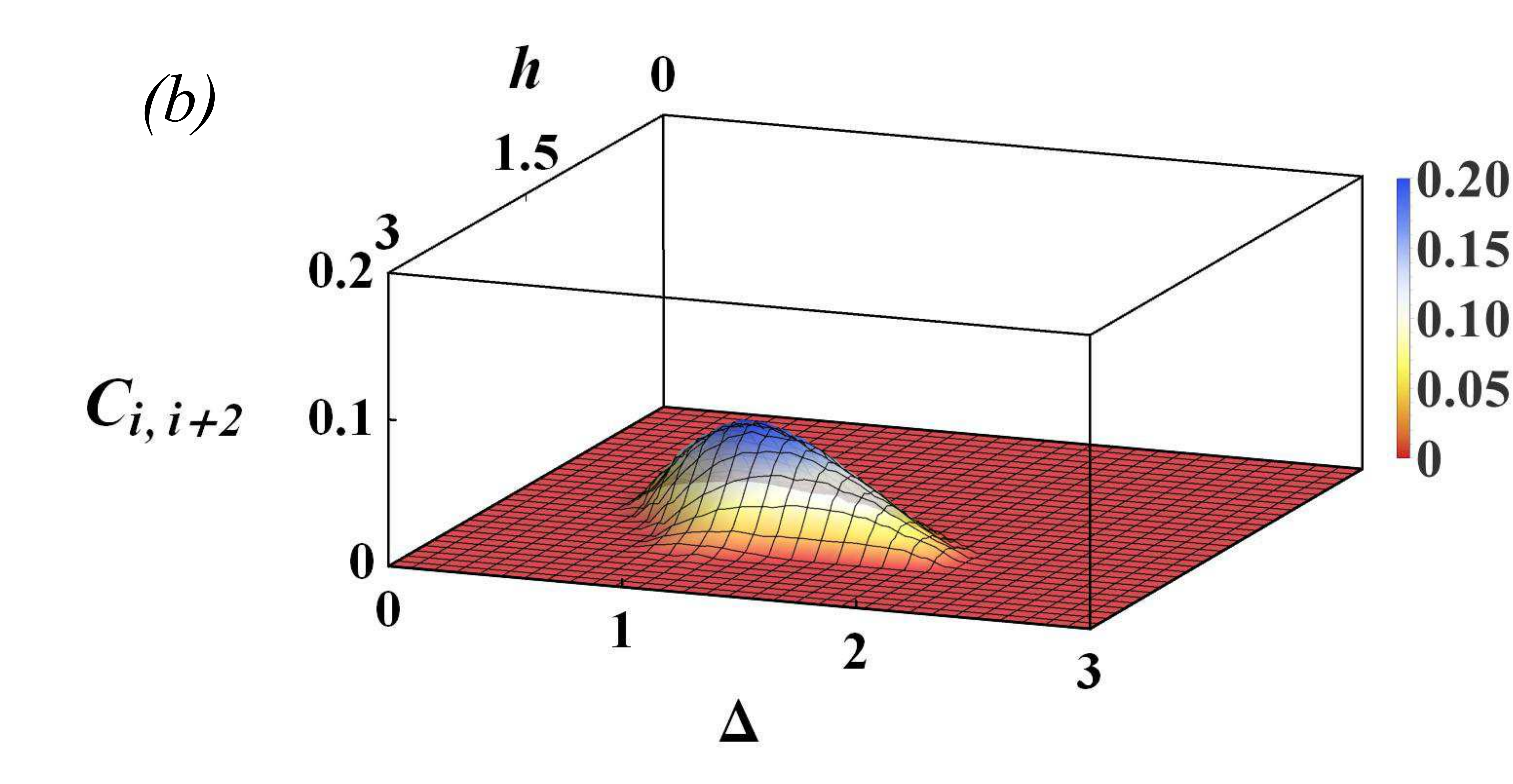}
\includegraphics[width=0.33\linewidth]{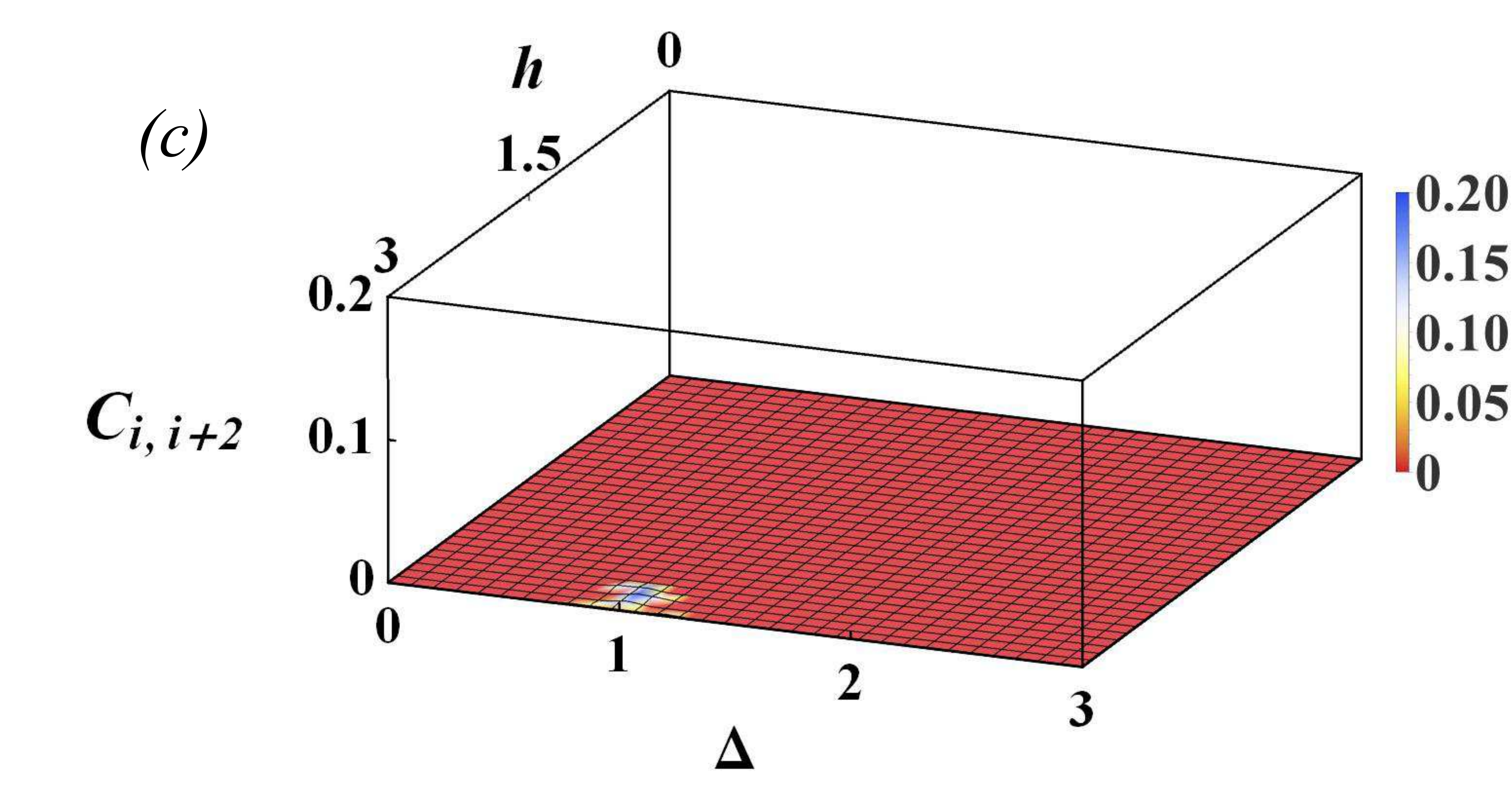}}
\caption{(Color online.) Three-dimension panorama of the entanglement as a function
of magnetic field and anisotropy $\Delta$ at $T=0.1$, between the
(a) first (b) second nearest neighbors spins. (c) Three-dimension view of the entanglement as a function
of $h$ and $\Delta$ between the second nearest neighbors spins at $T=0.2$.}
\label{Fig5}
\end{figure*}
%
%
\begin{figure*}
\centerline{
\includegraphics[width=0.33\linewidth]{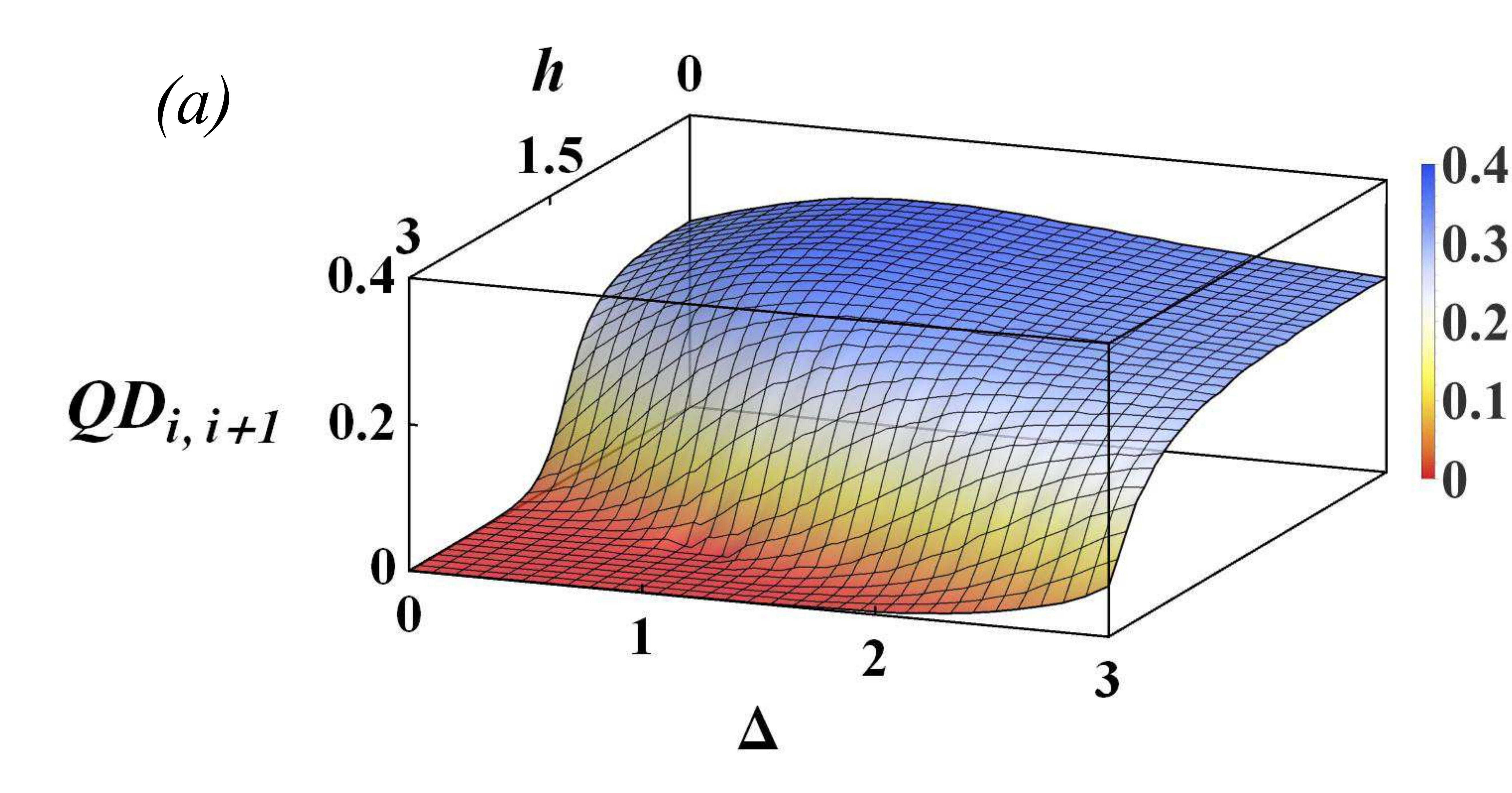}
\includegraphics[width=0.33\linewidth]{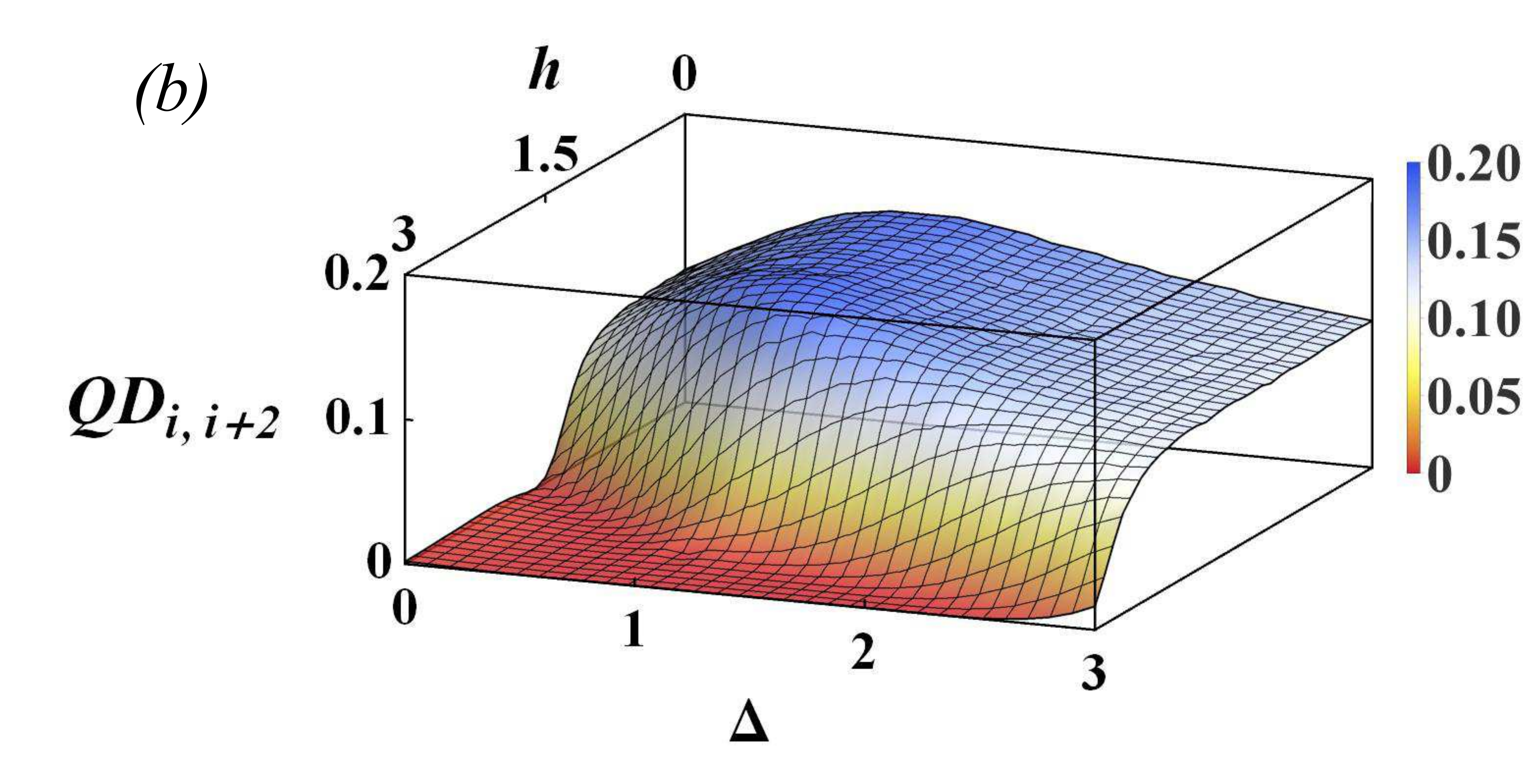}
\includegraphics[width=0.33\linewidth]{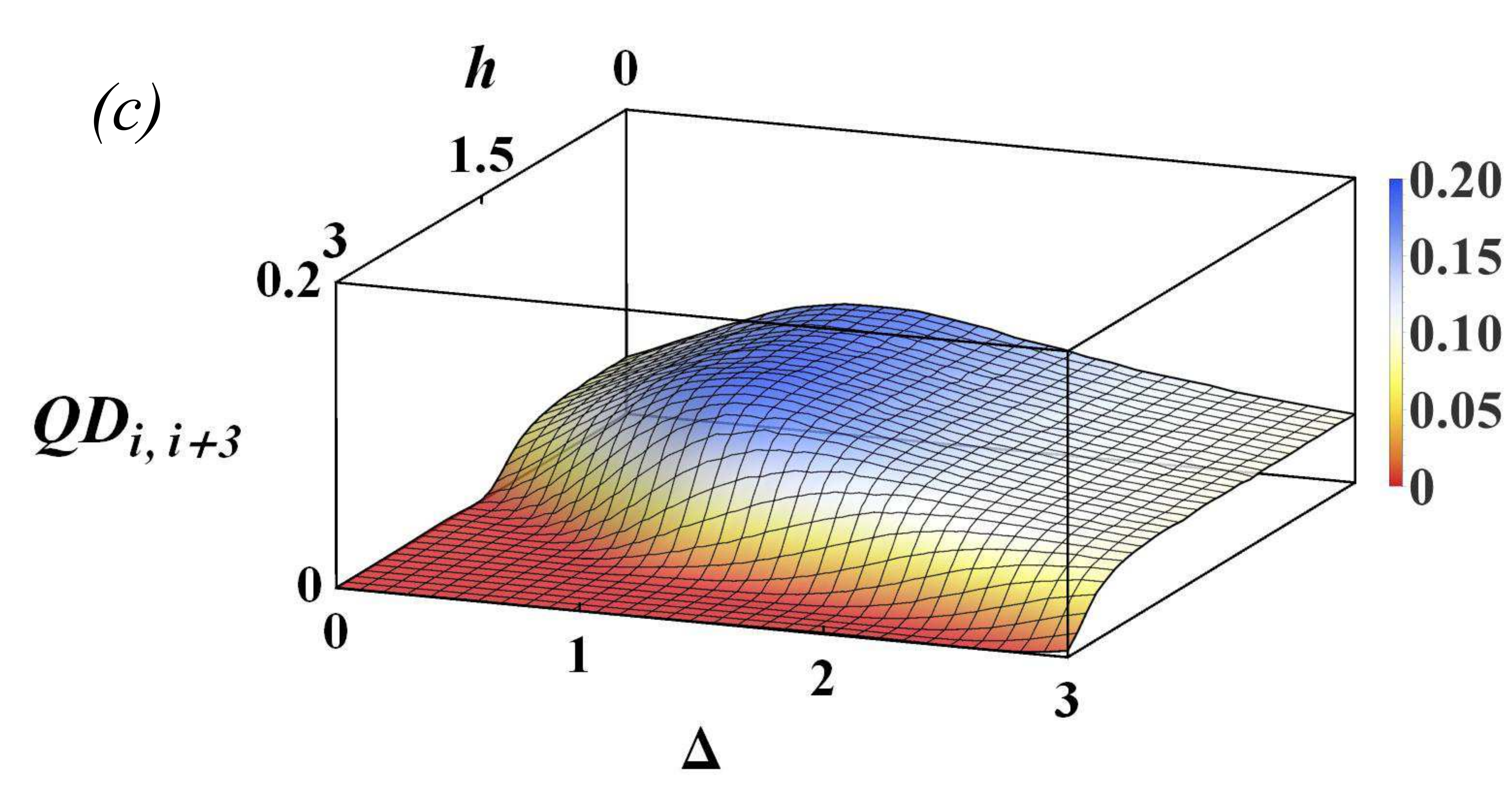}}
 \caption{(Color online.) Three-dimension view of quantum discord as a function
of magnetic field and anisotropy $\Delta$ at $T=0.1$, between the
(a) first (b) second and (c) third nearest neighbors spins.}
 \label{Fig6}
\end{figure*}
%
As seen in Fig. \ref{Fig3}(a), the entanglement between the 1st spin pairs remains finite up to the factorized
field ($h_{f}=J\sqrt{2 (1+\Delta)}$) and it clearly vanishes as expected for a factorized state where the ground state
of the chain is exactly separable. It should be mentioned that
in Refs. [\onlinecite{Roscilde04, Amico12}] entanglement between the 1st neighbor spins has been investigated
through the numerical approaches. It was shown that, for finite lattice size, entanglement shows a steep recovery
beyond the factorized field and it remain finite even in the saturated region $h>h_c$.
As expected, all QCs should be vanished in the saturated region. Consequently, the meanfield analytical approach
shows more accurate result than numerical method was used in Refs. [\onlinecite{Roscilde04, Amico12}].

As has been shown in Fig. \ref{Fig1}(a), only the nearest neighbor spin pairs are entangled when the transverse magnetic field is zero,
while the QD exists between spin pairs arbitrarily distant (Fig. \ref{Fig1}(b)).
It is seen from Fig. \ref{Fig3}(b) that, the second neighbor spin pairs become entangled in the presence of the transverse magnetic field for certain regions of parameter space. The 2nd neighbor spin pairs remain disentangled up to the desired critical entangled-field, $h_{c}^{E}<h_{f}$. In other words, there is a threshold transverse field $h_{c}^{E}$ above which the 2nd neighbor spins become entangled at zero temperature.
The entanglement which induced by an external magnetic field is known as the "magnetic entanglement" where reported for longitudinal magnetic fields \cite{Dillenschneider08, Khastehdel16, Yano05, Maziero10, Huang14}. It is worthwhile to mention that, the phase transition at $T=0$
is pinpointed by a global maximum of entanglement between both first and second neighbor spin pairs at the critical point $\Delta_{c}=1$.
However, at zero temperature the TF is incapable of creating the magnetic entanglement between spin pairs at distances beyond the next-nearest neighbors.
This behavior is in contrast with the case of XXZ in a longitudinal magnetic field \cite{Khastehdel16}.

To find more information about the magnetic entanglement region, we have calculated the critical entangled field for different
values of the anisotropy which result has been shown in Fig. \ref{Fig3}(c). As expected, the magnetic field which can create
entanglement between particles should have minimum strength at the critical point due to the divergence of correlation length.
As seen, the critical entangled field $h_{E}$ gives rise to the fingerprint of the quantum phase transition by displaying a minimum at the critical point $\Delta_c=1.0$.

Additionally, three dimensional panorama of the QD between the first, second and third neighbor spin pairs, at zero temperature, has been displayed
in Figs. \ref{Fig4}(a)-(c) versus $h$ and $\Delta$. As seen, the QD between the first, second and third neighbor spins
descends by the onset of TF and vanishes at $h_{QD}^{(|i-j|)}(\Delta)$.
The QD is also reserved between the 3rd neighbor spins in the presence of magnetic field where pairwise entanglement is absent.
We can see that, the QPT is characterized by a global maximum of QD at CP. Besides, the QD shows a cusp at the CP by increasing
the distance between the spin pairs Fig. (\ref{Fig4}(c)). This behavior intimates that the first derivative of the QD
between spin pairs beyond the next nearest neighbor distance is discontinuous at $\Delta_{c}=1$ and its second derivative
diverges at the CP.

To investigate the effect of temperature, we have plotted the entanglement between first and second nearest neighbor spins versus magnetic field
and anisotropy at $T=0.1$ in Figs. \ref{Fig5}(a)-(b). One can clearly see that, at low temperature, thermal entanglement behaves similar to the
zero temperature counterpart except that it maximum does not occurs at the critical point. Moreover, the entanglement decreases as temperature is increased and becomes disentangled gradually at high temperature (see Fig. \ref{Fig5}(c)). Our analysis also shows that, the critical entangled field ($h_{c}^{E}$) is not minimum in the presence of temperature. In other words, thermal entanglement is not able to detect the critical point even at low temperature.
It can be clearly seen that from Figs. \ref{Fig6}(a)-(c), TQD is more resistant to thermal effects than entanglement. At low temperature,
the QPT is still characterized by a global maximum of QD at CP while the cusp at the CP of TQD between far neighbors is eliminated by thermal fluctuation.
In contrast to the thermal entanglement, TQD is still a better estimator of the QCP in the presence of a transverse field.

%
\begin{figure*}
\centerline{
\includegraphics[width=0.33\linewidth]{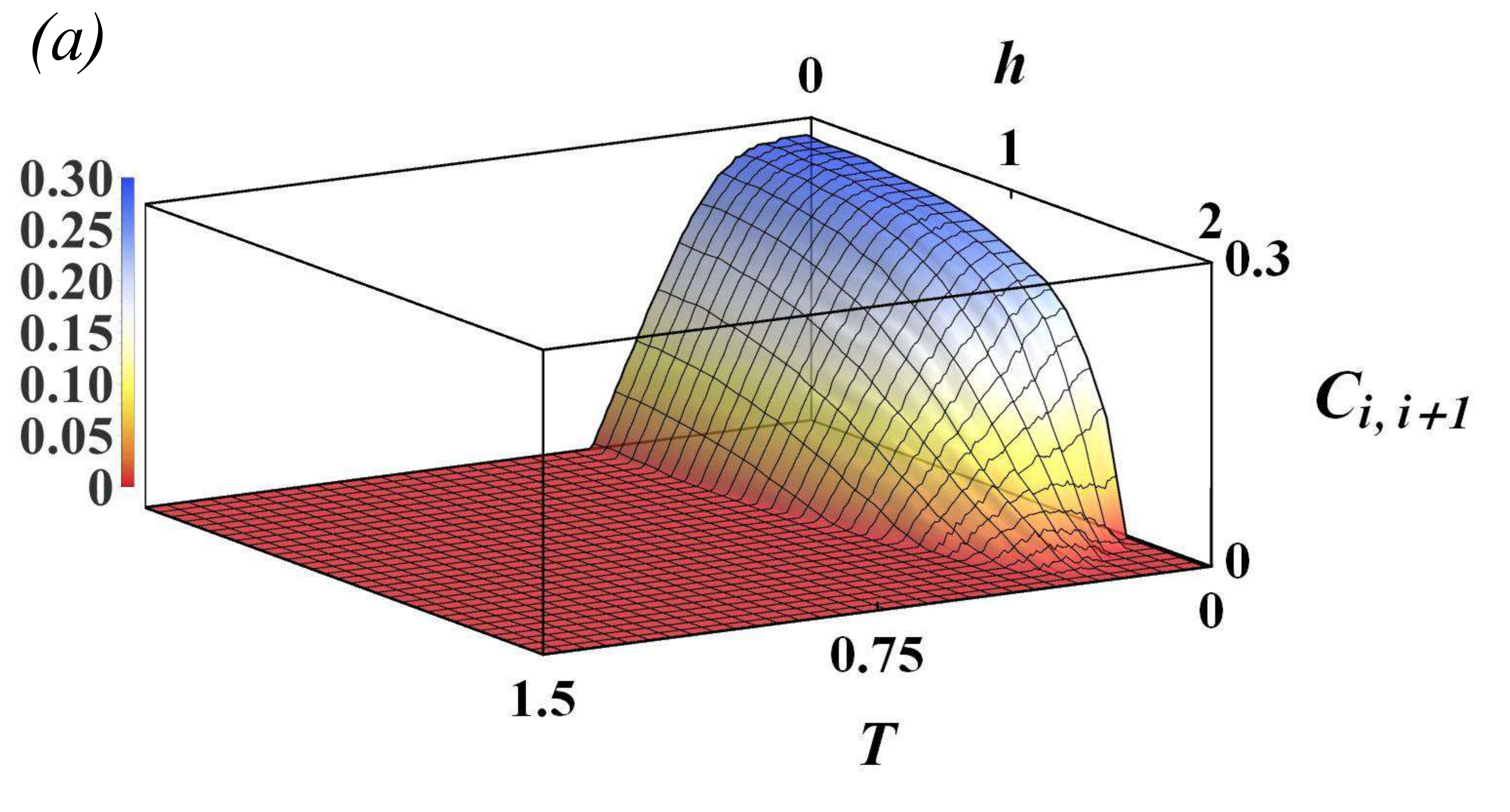}
\includegraphics[width=0.33\linewidth]{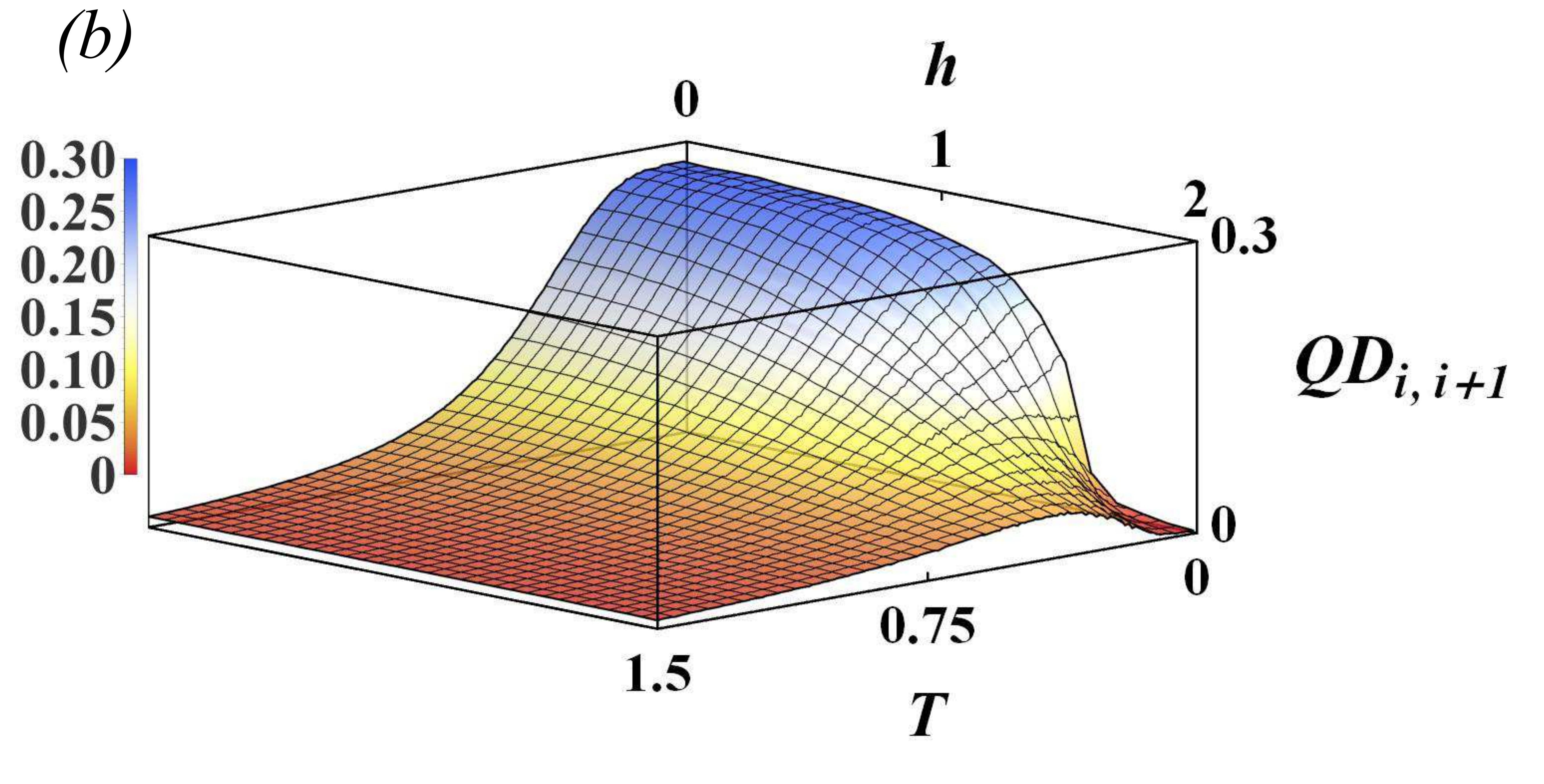}
\includegraphics[width=0.33\linewidth]{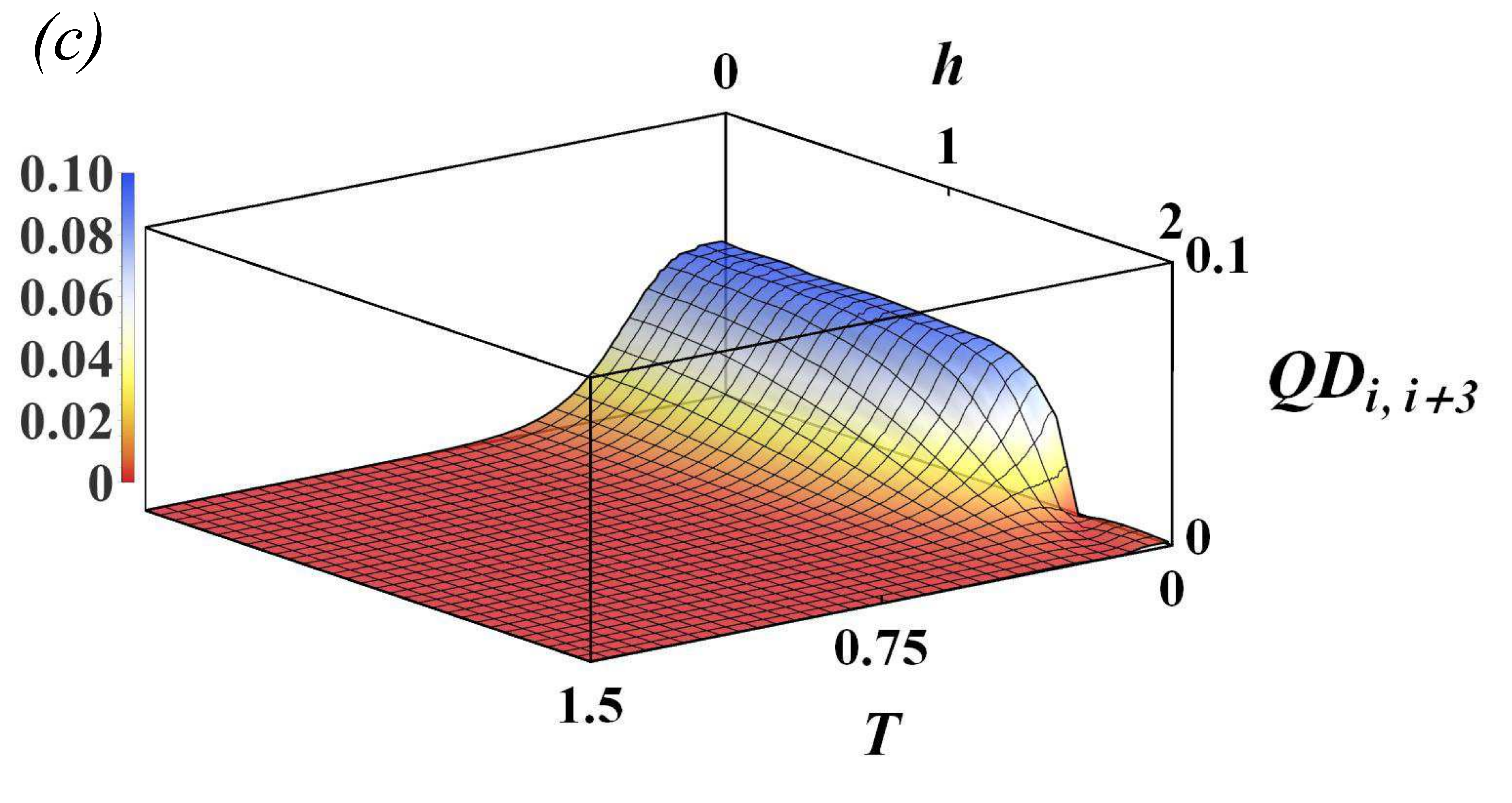}}
 \caption{(Color online.) (a) The thermal entanglement of formation between first neighbor spin pairs
(b) thermal quantum discord between first neighbor spin pairs and (c) thermal quantum discord between third neighbor spin pairs
as a function of magnetic field and temperature for $\Delta=0.5$.}
 \label{Fig7}
\end{figure*}
%

To complement our studies about the transverse field $XXZ$ model we present the thermal behavior of the quantum correlations in the presence of the
TF. It is necessary to mention that, in the saturation region, $h>h_c$, adding temperature creates neither the entanglement nor the QD.
In  Fig.~\ref{Fig7}(a)-(c), we have plotted the thermal entanglement and TQD as a function of the magnetic field and temperature for the anisotropy parameter $\Delta=0.5$. It is clear that, there is a critical values of field $h_{c}^{(E, D)}(\Delta)$ beyond which entanglement and QD disappear
at zero temperature and decline at finite temperature $h_{c}^{(E, D)}(\Delta, T)$. We have found also a critical temperature
$T_{c}^{(E, D)}(h, \Delta)$ after which entanglement and QD vanish, though there is a range of field (near to the factorized field) over which entanglement and QD can be increased by increasing the temperature.
Enhancing of entanglement and QD with temperature in the presence of magnetic field is a result of the fact that, the ground state tends to be less correlated than some low-lying excited states. So, correlated excited states are populated by increasing the temperature, in turn leading to the net effect of an increasing of entanglement and QD. This effect gets wiped out as the temperature gets too large. This behavior is similar to the behavior of entanglement reported in Ref. [\onlinecite{Arnesen}].


\section{Conclusion}\label{sec4}
We have studied the pairwise quantum correlations measured by the entanglement and the quantum discord
in the thermodynamic limit of the nonintegrable $XXZ$ spin-$1/2$ chain in a transverse magnetic field at
zero and finite temperatures. We have obtained analytical expressions for quantum correlations for
spin pairs at any distance. We have shown that the quantum discord between far neighbors is able to mark
the quantum phase transition, even for distances where pairwise entanglement is absent. This is the results of
the longer range of quantum correlation as quantified by quantum discord in comparison with the short-range behavior of pairwise
entanglement. Concerning the thermal effect onto quantum correlations, we have shown that thermal quantum discord between neighboring pairs displays
a strong distinctive behavior at the critical point that can be detected at finite temperature.
This significant property of thermal quantum discord is an important tool that can be easily applied to determine
quantum critical points of the systems which today's technology makes it virtually impossible to achieve the necessary $T$
below which quantum fluctuations dominate. Moreover, the thermal quantum discord behaves more robust than the thermal entanglement
as the temperature is increased. We have also shown that the transverse magnetic field creates the magnetic entanglement between the 2nd neighbor spins
in a narrow region under the factorized field. Remarkably, we show that quantum correlations can be increased with temperature in the presence of the magnetic field for certain regions of parameter space.

\section{acknowledgments}
The authors thank  Alireza Akbari and Utkarsh Mishra for valuable comments.
\vspace{0.3cm}
\bibliography{Ref}
\end{document}